\documentclass[reprint,aps,pre,superscriptaddress]{revtex4-2}
\usepackage{amsmath}
\usepackage{amsfonts}
\usepackage{amssymb}
\usepackage{xcolor}
\usepackage[dvipsnames]{xcolor}
\usepackage{soul}
\usepackage{graphicx}
\usepackage[colorlinks=true,linkcolor=blue,citecolor=blue,urlcolor=blue]{hyperref}
\usepackage[capitalize]{cleveref} 

\begin{document}

\title{Double descent: When do neural quantum states generalize?}

\author{M. Schuyler Moss}
\email{msmoss@uwaterloo.ca}
\affiliation{Department of Physics and Astronomy, University of Waterloo, Ontario, N2L 3G1, Canada}
\affiliation{Perimeter Institute for Theoretical Physics, Waterloo, Ontario, N2L 2Y5, Canada}

\author{Alev Orfi}
\affiliation{Center for Computational Quantum Physics, Flatiron Institute, 162 5th Avenue, New York, NY 10010, USA}
\affiliation{Center for Quantum Phenomena, Department of Physics, New York University, 726 Broadway, New York, New York 10003, USA}

\author{Christopher Roth}
\affiliation{Center for Computational Quantum Physics, Flatiron Institute, 162 5th Avenue, New York, NY 10010, USA}

\author{Anirvan M. Sengupta}
\affiliation{Department of Physics and Astronomy, Rutgers University, Piscataway, New Jersey 08854, USA}
\affiliation{Center for Computational Quantum Physics, Flatiron Institute, 162 5th Avenue, New York, NY 10010, USA}
\affiliation{Center for Computational Mathematics, Flatiron Institute, 162 5th Avenue, New York, NY 10010, USA}

\author{Antoine Georges}
\affiliation{Coll{\`e}ge de France, 11 place Marcelin Berthelot, 75005 Paris, France}
\affiliation{Center for Computational Quantum Physics, Flatiron Institute, 162 5th Avenue, New York, NY 10010, USA}
\affiliation{CPHT, CNRS, {\'E}cole Polytechnique, IP Paris, F-91128 Palaiseau, France}
\affiliation{DQMP, Universit{\'e} de Gen{\`e}ve, 24 quai Ernest Ansermet, CH-1211 Gen{\`e}ve, Suisse}

\author{Dries Sels}
\affiliation{Center for Computational Quantum Physics, Flatiron Institute, 162 5th Avenue, New York, NY 10010, USA}
\affiliation{Center for Quantum Phenomena, Department of Physics, New York University, 726 Broadway, New York, New York 10003, USA}

\author{Anna Dawid}
\affiliation{$\langle aQa^L\rangle$ Applied Quantum Algorithms -- Leiden Institute of Advanced Computer Science \\ \& Leiden Institute of Physics, Universiteit Leiden, The Netherlands}

\author{Agnes Valenti}
\affiliation{Center for Computational Quantum Physics, Flatiron Institute, 162 5th Avenue, New York, NY 10010, USA}

\date{\today}

\begin{abstract}

Neural quantum states (NQS) provide flexible and compact wavefunction parameterizations for numerical studies of quantum many-body physics. In particular, NQS aim to circumvent the exponential scaling of the Hilbert space by compressing quantum many-body wavefunctions with a tractable amount of parameters.
While inspired by deep learning, it remains unclear to what extent NQS share characteristics with neural networks used for standard machine learning tasks. We demonstrate that, in a simplified supervised setting, NQS exhibit the double descent phenomenon, a key feature of modern deep learning, where generalization worsens as network size increases before improving again in an overparameterized regime. Notably, we find the second descent to occur only for network sizes much larger than the Hilbert space dimension, i.e. network sizes that are out of reach for problems of practical interest. Within our setting, this observation places typical NQS in the underparameterized regime. 
We also observe that the optimal network size in the underparameterized regime depends on the number of unique training samples. While the double descent phenomenon does indeed translate to the NQS setting, potential practical consequences of our findings point more towards the need for symmetry-aware, physics-informed architecture design, rather than directly adopting machine learning heuristics.
\end{abstract}

\maketitle

\section{Introduction}
Variational approaches aim to circumvent the exponential cost of the quantum many-body problem by using an efficient parametrization of the wavefunction 
\cite{Becca_book_2017,Wu_Vscore_2024}. 
Traditionally, these parameterizations are guided by physical insight \cite{jastrow1955many-body-35c, bardeen1957theory, gutzwiller1963effect, anderson1987resonating-f7c, gros1989physics}.
Recently, a new paradigm has emerged that seeks more generic wavefunction ans\"atze. One such approach connects the ansatz structure to the entanglement in the system. For example, matrix product states and related tensor network extensions provide a systematically improvable variational 
framework \cite{white1992density, stoudenmire2012studying-14a, orus2014practical}, particularly effective in low-dimensional systems with area-law entanglement.
An even more generic and physics-agnostic approach draws from advances in machine learning. Artificial neural networks have repeatedly demonstrated their ability to efficiently process high-dimensional data and extract the underlying structure without prior knowledge of the problem.
Their representational power, guaranteed by universal approximation theorems \cite{cybenko1989, hornik1989, hornik1991}, makes the neural network parametrization of ground-state wavefunctions (called neural quantum states, NQS) \cite{carleo_solving_2017, carrasquilla2017machine} particularly appealing in regimes where other methods struggle.

While NQS were inspired by the success of deep learning, it remains an open question to what extent they share properties with neural networks used for standard machine learning tasks and whether deep learning heuristics translate to the quantum setting \cite{westerhout2020generalization,bukov2021learning-8c5, nutakki2025design-f7e, dash2025efficiency-013, barton2025interpretable-1c2, schurov2025learning-62d}.
For instance, an important observation in contemporary machine learning is that networks tend to have better predictive power when they are \textit{overparameterized}, i.e., when they have more parameters than needed to fit the training data. 
In this modern ``interpolating" regime \cite{nakkiran2021deep}, neural networks can fit the training data perfectly, achieving zero training error, yet they still generalize well to unseen (test) data. 
Indeed, many successes have followed from the modern intuition that ``bigger is better'' when it comes to network size. 
This stands in contrast to the classical regime of small, underparameterized models, where increasing the model size typically leads to overfitting due to the bias-variance trade-off~\cite{geman1992neural-bd1}.

\begin{figure}[t]
    \centering
    \includegraphics[width=\linewidth]{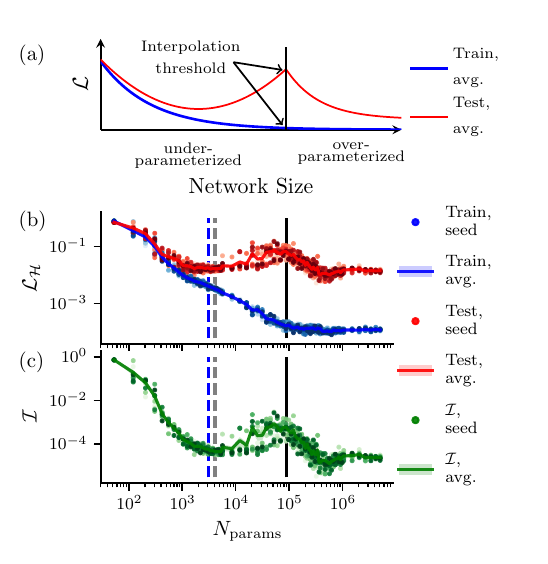}
    \caption{ 
    (a) A schematic showing the general features of double descent for deep neural networks~\cite{belkin_reconciling_2019}.
    (b) Training and test loss as a function of the number of network parameters when our NQS are trained on $\mathcal{D}_{\rm Train}^{\mathrm{top}\,75\%}$. Markers represent the loss for an individual trained network, and the solid lines represent the averages over ten random initializations. 
    (c) The infidelity between the trained NQS and the true ground state $\vert\Omega\rangle$.
    The black vertical line represents our estimate of the interpolation threshold. The gray dashed line (blue dashed line) indicates where the number of network parameters equals the size of the Hilbert space, $N_\mathrm{params}=2^N$ (the number of training configurations, $N_\mathrm{params}=75\% \times 2^N$).
    }
    \label{fig:main_DD}
\end{figure}

These two regimes, classical (underparameterized) and modern (overparameterized), are often connected by a characteristic “double descent” behavior as in~\cref{fig:main_DD}(a): Network performance initially degrades with increasing size, reaching the so-called interpolation threshold, before improving again in the overparameterized regime \cite{belkin_reconciling_2019, rocks2022memorizing-b5e}.
Double descent has been consistently observed across a wide range of machine learning tasks \cite{loog2020brief-99d, nakkiran2021deep, kempkes2025double-4d9} and its underlying mechanisms, albeit not fully understood, are explored in numerous works \cite{adlam2020understanding, liao2021random-62b, geiger2021landscape-15c, rocks2022memorizing-b5e, maloney2022solvable, Curth2023uturnondoubledescent, schaeffer2023double-46a, Gu2024unravelingdoubledescent, bach2024high}.
It is, however, unclear whether the double descent behavior, and its favorable implications, also hold when learning a quantum many-body wavefunction. In other words, does double descent emerge in NQS, and if so, can NQS benefit from this overparameterized regime?

In this work, we report the observation of the double descent phenomenon in NQS. We devise a setup that probes for the generalization properties of NQS by mirroring more conventional machine learning tasks. More concretely, we frame the learning of a quantum many-body wavefunction as a supervised learning task, partitioning the Hilbert space into a training and test dataset where the label for each spin configuration is the exact ground-state amplitude of the transverse-field Ising model (TFIM). Then, we minimize a mean-squared loss function to train fully connected, physics-agnostic neural networks on the training set. 
We find that the test loss exhibits a clear double descent behavior as the network size increases, with a peak at the interpolation threshold —  the position at which the training loss reaches its minimum value, see \cref{fig:main_DD}(b). Notably, this threshold lies well beyond the Hilbert space size, indicating that, under this supervised learning setup, {\it NQS operate in the underparameterized regime}. We stress that we observe this behavior despite the fact that we are learning the ground states of a simple Hamiltonian that is, in principle, exactly solvable. While we do not confirm that our observations hold for more complex Hamiltonians, there is no clear mechanism by which larger systems or richer interactions would would shift the interpolation threshold to smaller networks. Therefore, our results point to the need for physically-informed ans\"{a}tze \cite{vieijra2020restricted-3ac, vieijra2021many-body-27c,  Valenti2021,roth2021groupconvolutionalneuralnetworks, luo2021gauge-03b, robledo2022fermionic, luo2023gauge-invariant-6a8, machaczek2025neural, kufel2025approximately, lange2024simulating-0d5,chen2025neural-78f} rather than relying on the ``bigger is better'' heuristic.
We further find that the location of the first minimum in the test loss depends on the number of unique training configurations, highlighting an interplay between network size and sampling, in our setting. 
Finally, we examine properties of the underlying physical state represented by the trained NQS to probe the origin of the double descent peak, and identify how NQS overfit at the interpolation threshold. 
Through extensive additional experiments, summarized in \cref{app:additional_experiments}, we show that our results are robust to different types of training data, the various phases of the TFIM, and larger system sizes.

\section{Setup}
A neural quantum state (NQS)~\cite{medvidovic2024neural, lange2024architectures, dawid2025machine-7cd} is a neural-network-based parameterization of a (typically unnormalized) quantum many-body wavefunction. 
We consider systems with $N$ spin-$\frac{1}{2}$ degrees of freedom and networks with parameters $\theta$ that map each spin configuration $\vec{\sigma}$ (a $z$-basis state) to a real or complex wavefunction amplitude $\psi_{\theta}(\vec{\sigma})$. The learned wavefunction can be constructed as $\vert\Psi_\theta \rangle=  \mathcal{N}^{-1}\sum_{\left\{\vec{\sigma}\right\}}\psi_\theta(\vec{\sigma})\vert\vec{\sigma}\rangle$, where $\mathcal{N}$ is a normalization constant.
The goal is that $\vert\Psi_\theta\rangle$ is a compressed approximation of a target state $|\Omega\rangle$, where the number of parameters is far less than the Hilbert space size $2^N$. A typical target state is the ground state of a many-body Hamiltonian $\hat{H}$, which NQS can be trained to approximate by minimizing the variational energy, $E_\theta = \langle \Psi_\theta\vert \hat{H}\vert\Psi_\theta\rangle$.
Notably, this typical NQS learning task is considered ``unsupervised'', as the ``training data'', spin configurations sampled using Markov chain Monte Carlo from the NQS distribution $\vert\Psi_\theta\vert^2$, are unlabeled and change during optimization.

Here, we aim to probe for generalization properties of NQS, in a setup that 1) allows for a straightforward notion of \textit{test loss}, which measures how well a trained network performs on unseen data, and 2) isolates the network's generalization properties from potential effects that arise as a consequence of variational energy minimization.
To this end, we \textit{transform the standard NQS setup into a ``supervised'' learning task}, which mirrors more conventional machine learning tasks such as image recognition. In our supervised setup, each input (spin configuration) is paired with a known ``label'' (the corresponding exact ground-state amplitude). More concretely, for small system sizes, the Hamiltonian $\hat{H}$ can be exactly diagonalized, giving us access to all $2^N$ ground-state amplitudes $\left\{\Omega({\vec{\sigma}})\right\}$. We then partition the complete set of spin configurations $\left\{\vec{\sigma}\right\}$ and their corresponding amplitudes $\left\{\Omega({\vec{\sigma}})\right\}$
into a training set $\mathcal{D}_\mathrm{Train}$ and a test set $\mathcal{D}_\mathrm{Test}$. 
We consider the following types of datasets: 
\begin{itemize}
    \item $\mathcal{D}_{\rm Train}$$=\mathcal{D}_{\rm Train}^{\mathrm{top}\,X\%}$ consists of the X$\%$ of configurations in the Hilbert space with the largest exact ground-state amplitudes, i.e., $\Omega(\vec{\sigma})\geq \Omega(\vec{\sigma}^\prime)$ $\forall \vec{\sigma} \in \mathcal{D}_{\rm Train}^{\mathrm{top}\,X\%}$, $\vec{\sigma}^\prime \in \mathcal{D}_{\rm Test}^{\mathrm{top}\,X\%}$,
    \item $\mathcal{D}_{\rm Train}$$=\mathcal{D}_{\rm Train}^{\mathrm{unif.}\,75\%}$ consists of 75$\%$ of configurations in the Hilbert space drawn uniformly at random (with probability $p(\vec{\sigma}) = 1/2^N$), without replacement. We generate ten such $\mathcal{D}_{\rm Train}^{\mathrm{unif.}\,75\%}$ with different random seeds and ten corresponding test sets, where $\mathcal{D}_{\mathrm{Test}}^{\mathrm{unif.}\,75\%} = \left\{\vec{\sigma}\right\} \setminus \mathcal{D}_{\mathrm{Train}}^{\mathrm{unif}\,75\%}$,
    \item $\mathcal{D}_{\rm Train}=\mathcal{D}_\mathrm{Train}^{\mathrm{IS}\, 75\%}$ consists of the 75$\%\times2^N$ configurations, importance sampled (i.e. sampled with replacement) from the Born distribution corresponding to the true ground-state wavefunction $p_\Omega = \vert\langle\Omega\vert\Omega\rangle\vert^2$. We also generate ten such $\mathcal{D}_{\rm Train}^{\mathrm{IS}\,75\%}$ and corresponding $\mathcal{D}_{\rm Test}^{\mathrm{IS}\,75\%}$.
\end{itemize}
We provide more details about all of the training datasets considered in this work in \cref{app:data}.

Given a specific construction of a training and test set, we proceed to train the NQS on $\mathcal{D}_\mathrm{Train}$ to learn the mapping from spin configurations to wavefunction amplitudes. Specifically, since we consider ground states with non-negative, real-valued wavefunction amplitudes (see below), we minimize a loss function inspired by the Hellinger distance~\cite{Hellinger+1909+210+271,JeffreysHarold1946AIFf}:
\begin{align}
    \mathcal{L}_\mathcal{H}(\psi_\theta,\Omega) = \frac{1}{\sqrt{2}} \sqrt{
    \sum_{\vec{\sigma}\in \mathcal{D}_{\mathrm{Train}}}\big( \psi_\theta(\vec{\sigma}) - 
    \Omega(\vec{\sigma})\big)^2
    }.
    \label{eq:loss}
\end{align}
Note that $\psi_\theta(\vec{\sigma})$ is the unnormalized amplitude of the learned wavefunction.
We probe for double descent by assessing the generalization ability of the trained NQS as a function of total number of network parameters using the test loss, i.e., $\mathcal{L}_\mathcal{H}$ evaluated on $\mathcal{D}_\mathrm{Test}$.
Throughout this work, we use a three-layer feed-forward neural network as our NQS architecture, and control the total number of parameters by varying the width of the intermediate layers. We fixed the depth of our NQS because we observed that deeper networks did not provide significant performance gains and were more sensitive to hyperparameter choices. See \cref{app:Hellinger}, \cref{app:architecture}, and \cref{app:optimization} for more details about our loss function, architecture, and training, respectively. 

We perform the described experiments on the paradigmatic one-dimensional transverse-field Ising model (TFIM) with periodic boundary conditions. This Hamiltonian is exactly solvable, and therefore well-understood, and it is one of the standard benchmarks for NQS methods:
\begin{align}
    \hat{H} = \sum_{i=1}^{N-1} \hat{\sigma}^z_i\hat{\sigma}^z_{i+1} + \hat{\sigma}^z_N\hat{\sigma}^z_1- h \sum_i \hat{\sigma}^x_i.
    \label{eq:Hamiltonian}
\end{align}
The field strength $h$ controls a phase transition from a ferromagnet to a paramagnet with algebraically decaying correlations at the critical point $h=1$.
The ground state of the TFIM has real and non-negative amplitudes, which makes it a favorable test case for NQS, as no complex parameters are required to represent it.

\section{Results}
We use the setup described above to directly probe the generalization abilities of NQS. In what follows, we focus on our results for NQS trained to learn the ground state of the critical TFIM with transverse field $h=1$ and $N=12$. To support our main findings, we performed many additional experiments, all of which are summarized in \cref{app:additional_experiments}. 
In particular, we trained our NQS to learn ground states in different phases of the TFIM, namely the paramagnetic phase with $h=5$ and the ferromagnetic phase with $h=0.5$ (see \cref{app:phases}). We also examined NQS trained to learn the ground state of the critical TFIM for $N=16$ (see \cref{app:N16}).

\subsection{Double descent}
In~\cref{fig:main_DD}(b), the neural-network training and test loss are shown as a function of network size, for networks trained to represent the ground state of the TFIM at $h=1$ for $N=12$ spins. The NQS are trained on $\mathcal{D}_{\rm Train}^{\mathrm{top}\,75\%}$, which consists of the $75 \%$ of configurations in the Hilbert space {\it with the largest exact ground-state amplitudes}. This training data yields a simple learning problem, as $\mathcal{D}_{\rm Train}^{\mathrm{top}\,75\%}$ contains nearly all possible information about the target ground state.
We observe clear features of double descent: The test loss peaks at the interpolation threshold (marked with a black solid line), which corresponds to the smallest network size that fits the training data with the highest achieved accuracy (lowest achieved training loss). 
Notably, the observed interpolation threshold occurs at a parameter count that exceeds both the number of training configurations (blue dashed line) and the Hilbert space dimension (grey dashed line). This behavior is consistent across ten different random network initializations and suggests that, in this setting, NQS, which aim to compress the wavefunction representation using $N_{\rm params} \ll 2^N$, \textit{operate in the underparameterized regime.}

We further investigated the generalization ability of the networks using a physically meaningful measure of the quality of the NQS ground-state approximation, the infidelity between the trained NQS and the exact ground state $\mathcal{I} = 1-\vert\langle \Psi_\theta\vert\Omega\rangle\vert^2$. 
In \cref{fig:main_DD}(c), the behavior of the infidelity closely resembles the behavior of the test loss and spans several orders of magnitude, confirming that the observed double descent behavior in the test loss of \cref{fig:main_DD}(b) is of direct relevance to the overall physical accuracy of the obtained ground-state approximation. 

\subsection{Dependence on dataset size}
We extend the above analysis for NQS trained on smaller subsets of the Hilbert space, specifically $\mathcal{D}_{\rm Train}^{\mathrm{top}\,50\%}$ and $\mathcal{D}_{\rm Train}^{\mathrm{top}\,25\%}$, which contain the top $50 \%$ and $25 \%$ of configurations with the largest amplitudes, respectively (see \cref{app:less_data}). Interestingly, we find that the first minimum in the test loss, which lies within the underparameterized regime, occurs when the number of network parameters is approximately equal to the number of training configurations. This observation is also consistent with experiments where we trained our NQS using $\mathcal{D}_{\rm Train}^{\mathrm{IS}\,75\%}$, datasets created by importance sampling from the Born distribution corresponding to the true ground state wavefunction, and which better mimic the data used during variational NQS optimization. In this case, the minimum appears when the number of parameters matches the number of \textit{unique} training configurations (see \cref{app:IS_data}).
Our observations indicate an interplay between the network size and the number of (unique) training samples in our supervised learning setup. This interplay may persist and should be explored further in the more realistic variational NQS setting.

\begin{figure}
    \centering
    \includegraphics[width=\linewidth]{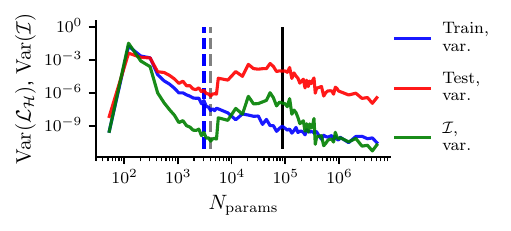}
    \caption{
    The variance of the training and test loss and the infidelities presented in \cref{fig:main_DD} as a function of the number of network parameters.
    The variance is taken across ten random initializations for each network size.
    The vertical lines follow the same convention as in \cref{fig:main_DD}. }
    \label{fig:DD_vars}
\end{figure}

\subsection{Rugged test loss landscape}
The overparameterized regime of deep neural networks is associated not only with improved generalization but also with a smoother loss landscape and a larger number of equivalent well-generalizing minima~\cite{Li2018visualizingloss, Garipov2018modeconnectivity, Simsek2021geometrysymmetry, Karhadkar2024favorablelandscape}.
Here, we test whether this holds for our NQS by analyzing the ruggedness of the loss landscape. Specifically, we quantify whether different random network parameter initializations lead to distinct minima in terms of training loss and other observables.
The more ``rugged'' the landscape is, the higher the variance of the obtained loss should be across different random seeds $s$, 
\begin{equation}
    \mathrm{Var}(\mathcal{L_{\mathcal{H}}}) = \frac{1}{S - 1} \sum_{s=1}^S \left( \mathcal{L_{\mathcal{H}}}^{(s)} - \bar{\mathcal{L}}_{\mathcal{H}} \right)^2.
\end{equation}

\Cref{fig:DD_vars} shows the variance of the quantities presented in \cref{fig:main_DD}. The variance of the training loss decreases monotonically with the number of network parameters, indicating that different initializations converge to the same training loss with increasing reliability. In contrast, the variance of the test loss and infidelity exhibit double descent behavior, peaking around the interpolation threshold identified in \cref{fig:main_DD}. This non-monotonic behavior indicates that the loss landscape around the interpolation threshold is highly degenerate, in which many parameter configurations achieve identical training loss. However, the degenerate minima correspond to functionally distinct solutions with varying generalization abilities, thus forming a rugged test loss landscape. In the overparameterized regime, the degenerate training loss minima encode increasingly similar solutions, as indicated by the second decrease in test loss and infidelity variances after the interpolation threshold. Thus, our overparameterized NQS induce a smoother test loss landscape. Our observation is consistent with the jamming perspective of the interpolation threshold in classical machine learning \cite{geiger2019jamming-23f}.

\subsection{Dependence on data composition} 
The choice of training data exposes the network to different physical features of the learning problem, affecting its generalization abilities.  
The results presented in \cref{fig:main_DD,fig:DD_vars} were obtained from NQS trained on $\mathcal{D}_{\rm Train}^{\mathrm{top}\,75\%}$, which is composed of configurations with the largest amplitudes of the exact ground-state wavefunction. 
To assess how generalization depends on data composition, we explore a contrasting scenario where NQS are trained on $\mathcal{D}_{\rm Train}^{\mathrm{unif.}\,75\%}$, which consists of spin configurations drawn uniformly at random. This training dataset resembles samples drawn from a randomly initialized NQS. We consider ten random data splittings, each of which produces a distinct training and test set.

\begin{figure}
    \centering
    \includegraphics[width=\linewidth]{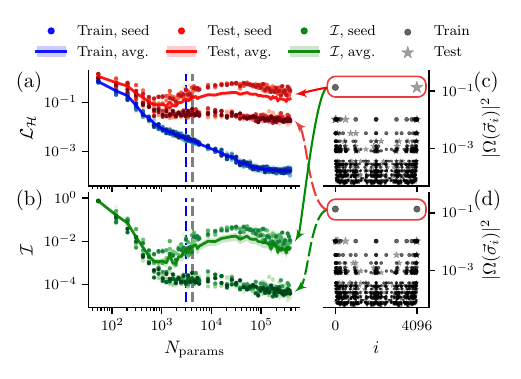}
    \caption{
    (a) Test and training loss for NQS trained on uniformly sampled training data $\mathcal{D}_{\rm Train}^{\mathrm{unif.}\,75\%}$. Markers represent the loss for an individual trained network, and the solid lines represent the averages over ten random initializations and datasets. 
    (b) Infidelity between the trained NQS and the true ground state $\vert\Omega\rangle$.
    The vertical lines follow the same convention as in \cref{fig:main_DD}.
    Panels (c) and (d) show the largest squared wavefunction amplitudes of the exact ground state, with dots and stars indicating the training and test configurations, respectively.
    These two dataset splittings exemplify the feature in the training data that leads to the two types of behavior in the test loss and infidelity. 
    In (c), only one of the two highest-probability configurations is in $\mathcal{D}_{\rm Train}^{\mathrm{unif.}\,75\%}$; in (d), both configurations are in $\mathcal{D}_{\rm Train}^{\mathrm{unif.}\,75\%}$.}    
    \label{fig:random_DD}
\end{figure}

As shown in \cref{fig:random_DD}(a), NQS trained on $\mathcal{D}_{\rm Train}^{\mathrm{top}\,75\%}$ still exhibit double descent behavior. The test loss, averaged over networks trained on the ten different datasets  (red solid line), shows a clear peak, though the peak is less pronounced and shifted to smaller network sizes than what is seen in \cref{fig:main_DD}. The infidelity between our trained NQS and the true ground state wavefunction, shown in \cref{fig:random_DD}(b), again confirms that the double descent behavior in the test loss has physical relevance. 

Interestingly, the test losses and infidelities corresponding to different datasets now separate into two distinct behaviors, with one group consistently demonstrating better generalization, i.e., smaller test loss and infidelity, than the other group.
This bifurcation reveals a strong dependence of the network's generalization abilities on the specific configurations seen during training. 
For the critical TFIM ($h=1$), the spin configurations with all spins up ($\vec{\sigma} =\,\uparrow\uparrow\dots\uparrow$) and all spins down ($\vec{\sigma} =\,\downarrow\downarrow\dots\downarrow$) have large ground-state amplitudes which dominate the Born distribution. All of the random datasets generated for these experiments contained one or both of these configurations and exemplary splittings are shown in \cref{fig:random_DD}(c) and (d), respectively. 
The NQS that consistently achieve lower test loss and infidelity with the true ground state wavefunction are those that were trained on datasets containing both high-probability configurations. 

\subsection{Exploring generalization}
To gain insight into the origin of the double descent peak and better understand how NQS overfit at the interpolation threshold, we analyze the trained networks by taking advantage of our knowledge of physical properties of the target ground state. 
First, we probe if the network correctly learns a generic feature of any physical quantum many-body state: the state's {\it normalization constant} $\mathcal{N}$. The amplitudes of the target ground state have a strong hierarchical structure in their magnitudes (see Fig.~\ref{fig:random_DD}), and $\mathcal{N}$ reveals whether a trained NQS systematically overestimates or underestimates the amplitudes of unseen test configurations. 

Second, we consider a physical property more specific to the TFIM. $\mathbb{Z}_2$ {\it parity} symmetry, namely $\vec{\sigma} \to \mathcal{P}\vec{\sigma}$, where the operator $\mathcal{P}$ flips all spins in $\vec{\sigma}$, plays a central role in the various phases of the TFIM. This symmetry is present in the paramagnetic phase and spontaneously broken in the ferromagnetic phase. At the critical point for a finite-size system, which we consider here, the ground state wavefunction also obeys this parity symmetry. Learning this symmetry would allow an NQS to generalize the target amplitude from one spin-configuration to its ``parity partner''. Probing the learned parity symmetry thus constitutes an interpretable measure of some of the network's generalization properties. We define and measure the parity error 
\begin{equation}\label{eq:parity_error}
    \epsilon_{\mathrm{parity}}= \frac{1}{|\mathcal{D}_{\mathrm{parity}}|}\sum_{\mathcal{D}_{\mathrm{parity}}} \left\vert1-  \frac{\psi_{\theta}(\mathcal{P}\vec{\sigma})}{\psi_{\theta}(\vec{\sigma})}\right\vert, 
\end{equation}
which measures how well the learned wavefunction respects the ground state’s parity symmetry for chosen configurations. 
Here, we consider parity-symmetric pairs where both $\vec{\sigma}$ and $\mathcal{P}\vec{\sigma}$ are in the test dataset, namely $\mathcal{D}_{\mathrm{parity}} = \left\{ \vec{\sigma} \;\middle|\; \vec{\sigma}, \mathcal{P}\vec{\sigma} \in \mathcal{D}_{\mathrm{Test}} \right\}
$.

\begin{figure}
    \centering
    \includegraphics[width=1\linewidth]{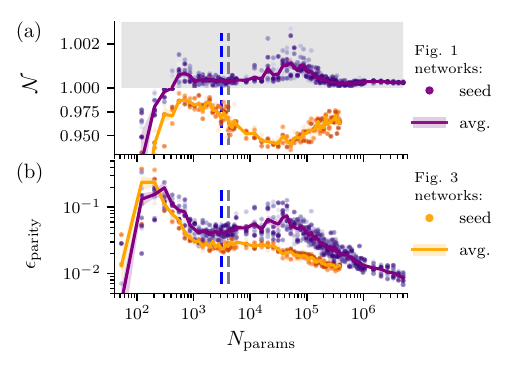}
    \caption{(a) The normalization constant $\mathcal{N}$ and
    (b) the parity error $\epsilon_{\rm parity}$ for NQS trained on different training datasets. Purple markers show these metrics for NQS trained on $\mathcal{D}_{\rm Train}^{\mathrm{top}\,75\%}$. Orange markers show these metrics for NQS trained on the $\mathcal{D}_{\rm Train}^{\mathrm{unif.}\,75\%}$ datasets which contain only a single high-probability configuration, e.g. shown in \cref{fig:random_DD}(c). Note the difference in y-axis scale above and below $\mathcal{N}=1$ in (a), marked by the shaded region. Markers, solid lines, and the vertical lines follow the same convention as in \cref{fig:main_DD}.}
    \label{fig:parity_norm}
\end{figure}

In \cref{fig:parity_norm}, we show the normalization constant and parity error for NQS trained on two types of training data: (i) $\mathcal{D}_{\rm Train}^{\mathrm{top}\,75\%}$ (used in \cref{fig:main_DD}, shown here in purple) and (ii) $\mathcal{D}_{\rm Train}^{\mathrm{unif.}\,75\%}$ (from \cref{fig:random_DD}, here in orange). For the latter, we focus only on the random datasets that produced more pronounced double descent, as in \cref{fig:random_DD}(c).

As shown in \cref{fig:parity_norm}(a), the normalization constant behaves oppositely for NQS trained on the two types of data near the interpolation threshold. For training data ordered by amplitude, $\mathcal{D}_{\rm Train}^{\mathrm{top}\,75\%}$, test configurations have small amplitudes by construction. When the network overfits to the training data, it tends to overestimate the amplitudes of configurations in the test set, resulting in $\mathcal{N}>1$ near the interpolation threshold. In contrast, for the uniformly split datasets considered here, namely $\mathcal{D}_{\rm Train}^{\mathrm{unif.}\,75\%}$ containing only one high-probability configuration, the network underestimates the amplitude of the high-probability configuration in the test set, yielding $\mathcal{N}<1$ for the networks that overfit, i.e. the networks corresponding to the peak in the test loss in \cref{fig:random_DD}(a).
The normalization constant learned by our trained NQS, manifests itself in the accuracy of various correlation functions, as explored in \cref{app:correlations}.

A similar contrast between the behavior of NQS trained on different datasets appears in the parity error. Networks trained on amplitude-ordered data, $\mathcal{D}_{\rm Train}^{\mathrm{top}\,75\%}$, exhibit double descent behavior in $\epsilon_\mathrm{parity}$. In this case, the peak in $\epsilon_\mathrm{parity}$ occurs at the same location as the peak in the test loss in \cref{fig:main_DD}(b), indicating that intermediate-sized NQS do not learn the parity symmetry when they overfit to the training data.
In contrast, networks trained on uniformly split data, $\mathcal{D}_{\rm Train}^{\mathrm{unif.}\,75\%}$, show steadily improving parity learning with increasing network size. For these NQS, $\epsilon_\mathrm{parity}$ monotonically decreases, meaning networks do learn the parity symmetry when overfitting to the training data. 
We note, however, that the parity error defined in Eq.~\ref{eq:parity_error} measures how well parity is learned {\it on average}. Therefore, a small $\epsilon_\mathrm{parity}$ does not imply that the parity symmetry is equally well-learned for all configurations. In fact, we find that the parity error between symmetric configurations $\vec{\sigma}$ and $\mathcal{P}\vec{\sigma}$ depends on the magnitude of their amplitudes $\vert\Omega(\vec{\sigma})\vert = \vert\Omega(\mathcal{P}\vec{\sigma})\vert$. When examining the parity error for specific configurations, we observe double descent behavior in $\epsilon_\mathrm{parity}$, but only for the few configurations with largest associated amplitudes. \Cref{app:parity} details this additional analysis. 
In sum, the parity symmetry is continuously learned for increasing network size, yet the learned symmetry is sacrificed or ignored for some configurations with large ground-state amplitudes when the NQS overfit.

Although both data splits show double descent in the test loss, the reasons for the test loss peak and the physical features captured by the networks differ with the structure of the training data. 

\section{Discussion and outlook}
In this work, we have demonstrated the double descent phenomenon in the context of learning a quantum many-body wavefunction in a supervised setting. We identified two regimes, underparameterized and overparameterized, separated by a peak in test error, which is consistently observed across various types of training data. Furthermore, our analysis reveals that this peak depends on the training data structure and can be associated with a rugged loss landscape.

Detailed studies of simple models in the classical machine learning setting~ \cite{liao2021random-62b,rocks2022memorizing-b5e,maloney2022solvable,bach2024high}  suggest that double descent arises due to the overparameterized model's ability to access new predictive features and the intrinsic regularization associated with the training procedure. For more complex learning tasks~\cite{rocks2022memorizing-b5e, schaeffer2023double-46a}, however, there is no consensus about the underlying reasons behind the strong generalization ability of overparameterized networks. Our investigations into the physical features of the learned ground-state wavefunctions shed light on how the networks fail to generalize at the interpolation threshold. Nevertheless, the generalizing patterns our networks learn in the overparameterized regime remain an open question.

In our experiments, we observed that the second descent in the test loss appears for network sizes far exceeding the Hilbert space dimension. This implies that, in the desirable setting where both the number of training configurations and the number of parameters network parameters are much smaller than the Hilbert space size, our NQS operate in the underparameterized regime. Consequently, the common machine learning heuristic ``bigger is better'' does not directly apply, since increasing the network size towards the interpolation threshold can lead to poor generalization. Notably, this behavior arises even when learning ground states of the simple, exactly solvable Hamiltonian considered here. While we cannot claim that the same behavior will hold for more complex Hamiltonians, we see no mechanism by which the interpolation threshold would shift to smaller networks for larger systems or richer interactions.  This points to the need for careful selection of NQS architecture (especially size). In our setting, we also find that the optimal network size, in the desired regime where $N_\text{params}\ll2^N$, is dependent on the training data, as the number of parameters at which the network generalizes best in the underparameterized regime correlates with the number of unique training configurations. This finding suggests an interplay between the network size and the number of samples, which should be explored further in the more realistic variational setting.

Ultimately, our supervised setup differs from the standard variational approach used in typical NQS ground-state searches, which involve energy minimization via stochastic reconfiguration, and samples drawn from the current wavefunction approximation. 
In contrast, we considered a supervised learning protocol where training data is fixed and not drawn from the current wavefunction approximation, which allowed us to remove effects of sampling from the already complex learning problem.
While any of our findings concerning network expressivity and the complexity of the target wavefunction remain relevant, effects arising from the training landscape may differ in the variational setting. For instance, the origin of the double descent peak differs when networks are trained on different datasets, suggesting that such effects are specific to the training setting. Understanding how these phenomena manifest in more practical NQS setups remains an important direction for future work.

\section{Code availability}
Our code relies on Jax~\cite{jax2018github}, NetKet~\cite{netket3:2022}, NumPy~\cite{harris2020array}, and Matplotlib~\cite{Hunter:2007}.
All of the code needed to reproduce our data can be found on GitHub~\cite{projectGit}. Some data is made available, in addition to the scripts used to produce our figures. The full datasets, which involve training quantities for every random initialization and dataset splitting, can be shared upon request.

\section{Acknowledgments}
We thank Matija Medvidović for useful discussions.
SM would like to acknowledge financial support from the Natural Sciences and Engineering Research Council of Canada (NSERC) and the Perimeter Institute. Research at Perimeter Institute is supported in part by the Government of Canada through the Department of Innovation, Science and Economic Development Canada and by the Province of Ontario through the Ministry of Economic Development, Job Creation and Trade.

AO, CR, AG, DS, and AV acknowledge support from the Flatiron Institute. The Flatiron Institute is a division of the Simons Foundation. Furthermore, SM and AO would like to thank the Flatiron Institute for providing the computational resources required for this work. 

AD acknowledges support from the
Dutch National Growth Fund (NGF), as part of the Quantum Delta NL programme. DS acknowledges support from AFOSR through grant FA9550-25-1-0067 and from NSF through grant 2118310.

\appendix

\section{Additional experiments}
\label{app:additional_experiments}
In the main text, we focused on our results for NQS trained to learn the ground state of the critical TFIM with transverse field $h=1$ and $N=12$. Here, we summarize additional experiments which show that our results are robust to different types of training data, the various phases of the TFIM, and larger system sizes.

\subsection{Observing double descent using less training data}
\label{app:less_data}
In this section, we provide further evidence that the position of the minimum test loss in the classical, underparameterized regime depends on the number of configurations in the training set, as suggested already by \cref{fig:main_DD}. Here, we test its robustness by systematically varying the number of training configurations.

\begin{figure}
    \centering
    \includegraphics[width=\linewidth]{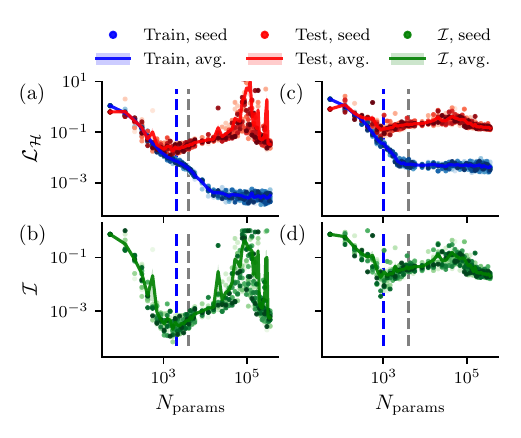}
    \caption{(a),(c) Training and test loss as a function of the number of network parameters when our NQS are trained on $\mathcal{D}_{\rm Train}^{\mathrm{top}\,50\%}$ and $\mathcal{D}_{\rm Train}^{\mathrm{top}\,25\%}$, respectively. (b),(d) The infidelity between the corresponding trained wavefunctions and the true ground state $\vert\Omega\rangle$.  Markers represent individual trained networks, and the solid lines represent the averages over ten random initializations. 
    The gray dashed line (blue dashed line) indicates where the number of network parameters equals the size of the Hilbert space, $N_\mathrm{params}=2^N$ (the number of training configurations, $N_\mathrm{params}=50\% \times 2^N$ in the first column and $N_\mathrm{params}=25\% \times 2^N$ in the second column).}
    \label{fig:DD_lessData}
\end{figure}

In \cref{fig:main_DD}, our NQS were trained on $\mathcal{D}_{\rm Train}^{\mathrm{top}\,75\%}$, which contains the 75\% of the Hilbert space with the largest ground-state amplitudes, corresponding to a training set size of $\vert\mathcal{D}_\mathrm{Train}^{\mathrm{top}\,75\%}\vert = 75\%
\times2^N$. We now consider smaller training sets, still composed of configurations with the largest ground-state amplitudes.
\Cref{fig:DD_lessData}(a)-(b) and (c)-(d) show the losses and infidelities of NQS when trained on the smaller datasets, namely $\mathcal{D}_{\rm Train}^{\mathrm{top}\,50\%}$ and $\mathcal{D}_{\rm Train}^{\mathrm{top}\,25\%}$ with $\vert\mathcal{D}_\mathrm{Train}^{\mathrm{top}\,50\%}\vert = 50\%\times2^N$ and $\vert\mathcal{D}_\mathrm{Train}^{\mathrm{top}\,25\%}\vert = 25\%\times2^N$, respectively. We observe that the first minimum in the test loss, which is in the underparameterized regime, indeed shifts to smaller network sizes. In both cases, the minimum aligns with the number of training configurations (indicated by the blue dashed line). We also continue to see clear double descent behavior across the different training set sizes.

We note that for NQS trained on the smaller training sets, the interpolation threshold is less well-defined. Here, the training loss reaches its minimum value at smaller network sizes than where the test loss exhibits the double descent peak, in contrast to the behavior observed in \cref{fig:main_DD}. Moreover, the separation between training loss minimum and test loss peak is larger for NQS trained on $\mathcal{D}_{\rm Train}^{\mathrm{top}\,25\%}$ than for NQS trained on $\mathcal{D}_{\rm Train}^{\mathrm{top}\,50\%}$. Therefore, for much smaller training sets, the minimal training loss can be reached in the underparameterized regime without strong overfitting. However, NQS trained using less training data become less accurate. The infidelities shown in \cref{fig:DD_lessData}(d) are notably larger than those shown in \cref{fig:DD_lessData}(b), and the infidelities in both subfigures are larger than those shown in \cref{fig:main_DD} in the main text.  

Finally, we note that our observation linking the first minimum in test loss with the number of unique training configurations does not hold for the NQS trained on $\mathcal{D}_{\rm Train}^{\mathrm{unif.}\,75\%}$ datasets, shown in \cref{fig:random_DD}. In that case, the first minimum occurs for network sizes that are smaller than the training set size, $N_\mathrm{params}< |\mathcal{D}_{\rm Train}^{\mathrm{unif.}\,75\%}|$.
This deviation likely stems from the nature of the training data: Unlike in \cref{fig:main_DD} and \cref{fig:DD_lessData}, the uniformly random datasets include configurations with very small amplitudes, which contribute a very small amount to the training loss. Such low-probability configurations are less informative about the ground state, potentially weakening the observed correlation between the test loss minimum and the size of the training set.

\subsection{Importance-sampled data}
\label{app:IS_data}
To better mimic the data used during the variational training of NQS, we introduce a third type of training dataset $\mathcal{D}_\mathrm{Train}^{\mathrm{IS}\, 75\%}$: datasets generated by importance sampling from the Born distribution corresponding to the true ground-state wavefunction $p_\Omega = \vert\langle\Omega\vert\Omega\rangle\vert^2$. This amounts to sampling configurations from $p_\Omega$ \emph{with replacement}, meaning configurations may be repeated in the training set. For ten seeds, we draw $N_s = 75\%\times 2^N$ configurations, generating ten distinct datasets. We emphasize that these datasets are sampled directly from the true Born distribution and are therefore independent of the networks trained on them. Due to the peaked nature of the distribution, however, the training datasets contain only $728 \pm 16$ unique configurations on average (averaged over the 10 datasets). The remaining configurations are considered the test dataset, $\mathcal{D}_{\mathrm{Test}}^{\mathrm{IS}\,75\%} = \left\{\vec{\sigma}\right\} \setminus \mathcal{D}_{\mathrm{Train}}^{\mathrm{IS}\,75\%}$. For more details about the different training datasets employed in this work, see \cref{app:data}.

\begin{figure}
    \centering
    \includegraphics[width=\linewidth]{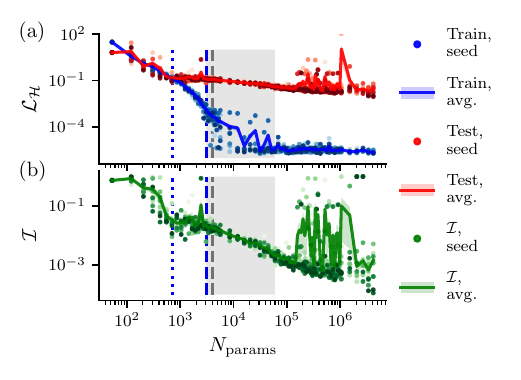}
    \caption{(a) Training and test loss as a function of the number of network parameters when our NQS are trained on $\mathcal{D}_\mathrm{Train}^{\mathrm{IS}\, 75\%}$. Markers represent the loss for an individual network trained on a given generated dataset, and the solid lines represent the averages over ten generated datasets. 
    (b) The infidelity between the trained wavefunctions and the true ground state $\vert\Omega\rangle$. 
    The gray dashed line (blue dashed line) indicates where the number of network parameters equals the size of the Hilbert space, $N_\mathrm{params}=2^N$ (the number of training configurations, $N_\mathrm{params}=75\% \times 2^N$). The blue dotted line indicates where the number of network parameters equals the average number of unique training configurations $N_\mathrm{params} = 728 \pm 16$ (averaged across the 10 datasets). The interpolation threshold lies in the shaded area, as that is where the training loss of many networks reaches its minimum value and the test loss begins its second descent.}
    \label{fig:DD_mc}
\end{figure}

When we train our NQS on $\mathcal{D}_\mathrm{Train}^{\mathrm{IS}\, 75\%}$, we do not observe double descent behavior in generalization metrics as clearly as in the other dataset splittings, as shown in \cref{fig:DD_mc}. In particular, the peak associated with double descent is less pronounced in the test loss shown in panel (a) than in the fidelity in panel (b). The interpolation threshold is also harder to identify than in \cref{fig:main_DD}. Networks trained on different training sets achieve the minimal training error at varying network sizes, which we mark with a shaded region. Nevertheless, this region seems to coincide with the location where the test loss peaks and begins its second descent.

Most importantly, we observe the first minimum of the test loss corresponds to the point where the number of parameters equals the number of unique training configurations, marked with a dotted blue line. This indicates that, in the underparameterized regime, the location of the test loss minimum is influenced by the number of \textit{unique} configurations in the training dataset, rather than the total number of configurations in the dataset.

Notably, the training loss is lower, and the test loss and infidelity are higher in this setting compared to most of our other experiments. This observation indicates stronger overfitting, despite this dataset being, in principle, more representative of the ground-state distribution than e.g. the uniform datasets considered in \cref{fig:random_DD}. The likely explanation for the lower training loss and the higher test loss lies in the small number of unique training configurations: It is easier for the network to memorize this dataset, but generalization becomes more challenging, as a larger portion of the configurations are excluded from the training.

Finally, we observe striking behavior for networks with $N_\mathrm{params} > 10^5$. Unlike in other experiments, where training remains stable for large networks, here we find a large variance in the test loss and the infidelity across different sampled datasets. This suggests that, in the regime where the training data includes only few unique configurations, large networks becomes harder to train, potentially due to increased difficulty in navigating the loss landscape.

\subsection{Double descent across different phases}
\label{app:phases}
\begin{figure*}
    \centering
    \includegraphics[width=0.9\linewidth]{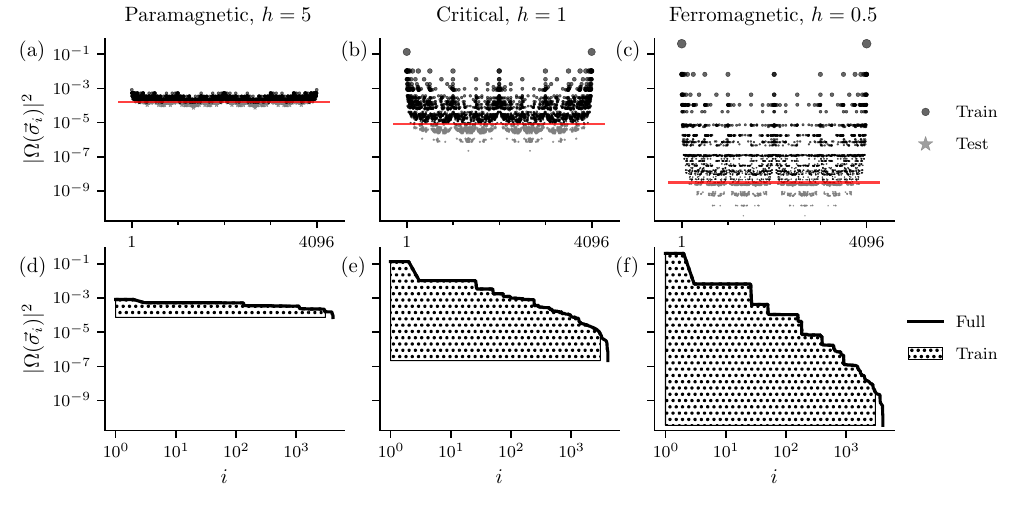}
    \caption{Panels (a)-(c) show the training dataset, $\mathcal{D}_{\rm Train}^{\mathrm{top}\,75\%}$, and corresponding test dataset, $\mathcal{D}_{\rm Test}^{\mathrm{top}\,75\%}$, for the three phases of the TFIM considered in this work: the paramagnetic phase with $h=5$, the critical phase with $h=1$, and the ferromagnetic phase with $h=0.5$. The red horizontal lines indicate the smallest Born-distributed probability $\vert\Omega(\vec{\sigma}_i)\vert^2$ in $\mathcal{D}_{\rm Train}^{\mathrm{top}\,75\%}$, or conversely, the largest probability in $\mathcal{D}_{\rm Test}^{\mathrm{top}\,75\%}$. Panels (d)-(e) show the probability density of $p_\Omega$ contained in each training dataset.}
    \label{fig:para_ferro_datasets}
\end{figure*}

The TFIM admits a paramagentic phase above and a ferromagnetic phase below the critical value of the transverse field, $h=1$. 
To further investigate the double descent phenomenon, we repeat the experiments used to produce Fig. 1 (b)-(c) in the main text for the TFIM deep in both of these phases. For the paramagnetic phase, we take $h=5$ and for the ferromagnetic phase, we take $h=0.5$. Recall that Fig. 1 (b)-(c) in the main text show results for the TFIM at criticality ($h=1)$. We train our NQS on $\mathcal{D}_\mathrm{Train}^{\mathrm{top}\,75\%}$, which contains the 75\% of configurations with the largest exact ground-state wavefunction amplitudes. The remaining configurations make up the test set, $\mathcal{D}_\mathrm{Test}^{\mathrm{top}\,75\%}$. Compared to the critical TFIM, the Born distribution corresponding to the true ground state in the paramagnetic phase is less peaked and closer to a uniform distribution. On the other hand, the Born distribution corresponding to the true ground state in the ferromagnetic phase is more peaked. In other words, the amplitudes of the ferromagnetic ground state span many more orders of magnitude and therefore, the configurations in $\mathcal{D}_\mathrm{Test}^{\mathrm{top}\,75\%}$ have very small Born-distributed probabilities. 
\Cref{fig:para_ferro_datasets} (a)-(c) display $\mathcal{D}_{\rm Train}^{\mathrm{top}\,75\%}$ and $\mathcal{D}_{\rm Test}^{\mathrm{top}\,75\%}$ for each of the phases of the TFIM. We also show the portion of the true Born distribution PDF that is contained in $\mathcal{D}_{\rm Train}^{\mathrm{top}\,75\%}$ in \cref{fig:para_ferro_datasets} (d)-(f).

\begin{figure*}
    \centering
    \includegraphics[width=0.8\linewidth]{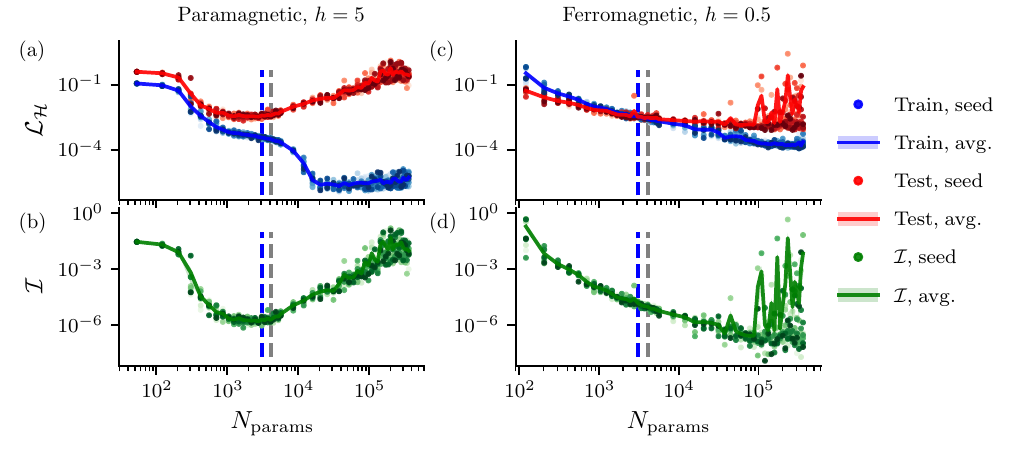}
    \caption{
    (a),(c) The final training and test loss achieved with our NQS as a function of the number of network parameters when the target ground state is in the ferromagnetic phase with transverse field $h=0.5$ and the paramagnetic phase with transverse field $h=5$, respectively. The networks are trained on the $\mathcal{D}_\mathrm{Train}^{\mathrm{top}\,75\%}$ datasets displayed in \cref{fig:para_ferro_datasets}(a) and (c), respectively. Markers represent the loss for an individual trained network, and the solid lines represent the averages over ten random initializations. 
    (b),(d) The infidelity between the trained wavefunctions from (a),(c) and the corresponding target ground state $\vert\Omega\rangle$. 
    The vertical lines follow the same convention as in \cref{fig:FE}.
    }
    \label{fig:para_ferro_dd}
\end{figure*}

\begin{figure*}
    \centering
    \includegraphics[width=0.8\linewidth]{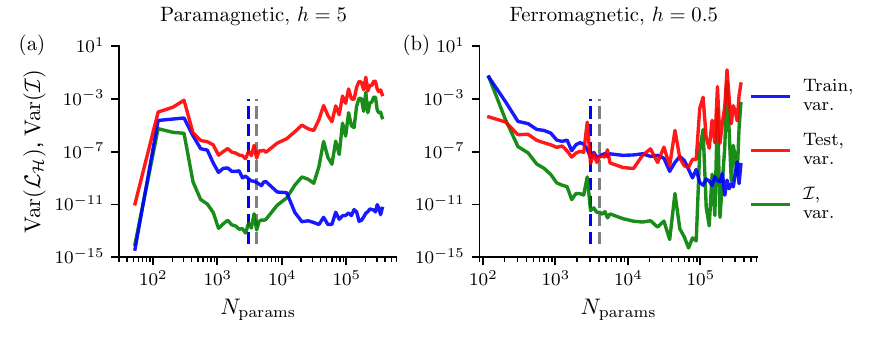}
    \caption{The variance of the training losses, test losses, and the infidelities presented in \cref{fig:para_ferro_dd} as a function of the number of network parameters. 
    Panel (a) corresponds to experiments shown in \cref{fig:para_ferro_dd} (a)-(b), where the target ground state is in the paramagnetic phase ($h=5$). Panel (b) corresponds to experiments shown in \cref{fig:para_ferro_dd} (c)-(d) where the target ground state is in the ferromagnetic phase ($h=0.5$).
    The variance is taken across the ten random initializations for each network size.
    The vertical lines follow the same convention as in \cref{fig:FE}.
    }
    \label{fig:para_ferro_vars}
\end{figure*}

Regardless of the phase of the target ground state and how that manifests in the training and test data, we observe the same qualitative double descent behavior as was reported in the main text. 
For both phases, \cref{fig:para_ferro_dd} shows similar behavior as Fig. 1 (b)-(c) in the main text. In particular, we see that increasing the network size systematically improves the train loss but leads to non-monotonic behavior in the test loss shown in (a),(c) and the infidelity with the target ground state shown in (b),(d). 
Furthermore, \cref{fig:para_ferro_vars} shows similar trends in the variance of the training and test loss and the infidelity as shown in Fig 2 in the main text. For both phases, the variance of the training loss remains small for large networks, while the variance of the test loss and infidelity again display non-monotonic behavior. The peak in the variance of the test loss and the infidelity suggests ruggedness in the test loss landscape because it indicates that our trained networks have reached different local minima with different generalization properties.

Interestingly, for the paramagnetic phase, the peak in the test loss is shifted to larger network sizes compared to the critical point (i.e. in Fig. 1 (b)-(c) in the main text, the peak is located around $N_{\rm params} \approx 10^{5}$), and it does not coincide with the minimum in training loss, which is shifted to smaller network sizes than for the critical point (similarly, see Fig. 1 (b)-(c) in the main text). Instead, the training loss reaches its minimum value for some network size $N_\mathrm{params}\approx2\times10^4$. Then, for a range of sizes $2\times10^4\leq N_\mathrm{params}\leq4\times10^5$, the networks continue to overfit to the data while achieving the same value of the training loss. This makes it difficult to mark the interpolation threshold, which formally occurs when the training loss reaches its minimum \emph{and} the test loss reaches its peak value. This behavior is similar to that in Fig. 5 in the main text and suggests that, for this data, there are multiple ways to overfit. Large networks can learn features that generalize increasingly worse, while maintaining the same minimal training error. For the ferromagnetic phase, it appears that there is a discernible interpolation threshold. The test loss reaches its peak at roughly the same number of parameters where the train loss reaches its minimum. Notably, the interpolation threshold in this case is shifted to larger network sizes compared to the critical point (Fig. 1 (b)-(c) in the main text).

As mentioned in \cref{app:optimization}, we focus only on networks with $W\leq432$ for these experiments due to the sensitivity of very wide networks during training. For our experiments where the target ground state is in the paramagnetic phase (\cref{fig:para_ferro_dd}(c)-(d) and \cref{fig:para_ferro_vars}(b)), the full double descent behavior is visible within this range of network sizes. It is clear that the test loss, infidelity, and their respective variances begin to decrease for the largest considered network sizes. For our experiments where the target ground state is in the ferromagnetic phase (\cref{fig:para_ferro_dd}(a)-(b) and \cref{fig:para_ferro_vars}(a)), we do not observe the second descent within the trainable network sizes. Nevertheless, 
the most important observation holds: Reaching the interpolation threshold, if it exists, requires a number of parameters that is larger than the Hilbert space size, which suggests that, in our experiments, NQS with $N_\textrm{params}\ll2^N$ operate in the underparameterized regime.

\subsection{Double descent for $N=16$ spins}
\label{app:N16}
\begin{figure*}
    \centering
    \includegraphics[width=0.8\linewidth]{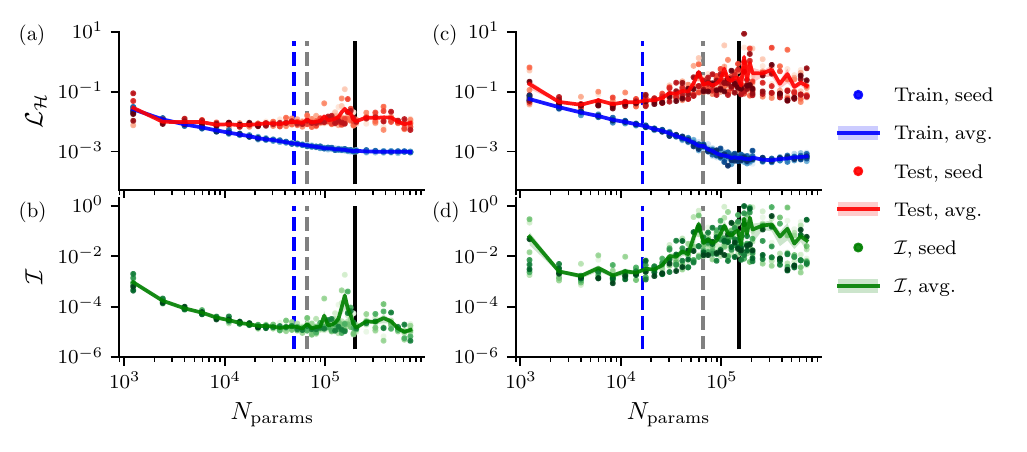}
    \caption{(a),(c) Training and test loss as a function of the number of network parameters when the NQS are trained on $\mathcal{D}_\mathrm{Train}^{\mathrm{top}\,75\%}$ and $\mathcal{D}_\mathrm{Train}^{\mathrm{top}\,25\%}$, respectively, for $N=16$ spins. (b),(d) The infidelity between the trained wavefunctions from (a),(c) and the corresponding target ground state $\vert\Omega\rangle$.  Markers represent individual trained networks, and the solid lines represent the averages over ten random initializations. 
    The black vertical lines represent our estimates of the interpolation thresholds for each set of experiments. The gray dashed lines (blue dashed lines) indicate where the number of network parameters equals the size of the Hilbert space, $N_\mathrm{params}=2^N$ (the number of training configurations, $N_\mathrm{params}=75\% \times 2^N$ in the first column and $N_\mathrm{params}=25\% \times 2^N$ in the second column).}
    \label{fig:N16}
\end{figure*}

Given that NQS are a promising candidate for compressing a quantum many-body wavefunction, it is common to employ them to study system sizes beyond the reach of exact methods. In order to understand how our results from the main text scale, we perform the same experiments for a larger system size. In particular, we consider the 1D TFIM with $N=16$ spins, where it is still possible to obtain the true ground-state wavefunction $\vert\Omega\rangle$, which we use to label the training data.

The size of the Hilbert space for a system with $N$ spin-$\frac{1}{2}$ degrees of freedom grows as $2^N$. In order to train NQS with a number of parameters $N_\mathrm{params}$ much larger than the size of the Hilbert space for $N=16$, we consider the network architecture described in Section IV with a depth of 4 instead of 3. We increase the batch size to $2^{10}$, and we use the exponentially decaying learning rate schedule for all widths. Other optimization details are consistent with what is described in Section II.

\Cref{fig:N16}(a) shows the training and test loss for NQS trained on $\mathcal{D}_\mathrm{Train}^{\mathrm{top}\,75\%}$, which contains the $75\%$ of configurations with the largest ground-state wavefunction amplitudes for the TFIM with $N=16$ and $h=1$. Similar behavior can be seen between the test loss in (a) and the infidelity between our trained NQS and the exact ground-state wavefunction in \cref{fig:N16}(b). The double descent behavior is subtle, but we identify our best estimate of the interpolation threshold, where the test loss peaks and the training loss reaches its minimal value.

Another important consequence of the exponentially larger Hilbert space is that the Born-distributed probabilities for configurations $p_\Omega(\vec{\sigma}) = \vert\Omega(\vec{\sigma})\vert^2$ span more orders of magnitude. For N=16, the most probable configuration has a probability on the order of $\mathcal{O}(10^{-1})$. The smallest probabilities, however, are less than $\mathcal{O}(10^{-8})$, and the average probability is on the order of $\mathcal{O}(10^{-5})$. This is considerably smaller than for $N=12$ (see \cref{fig:datasets}). The average probability of a configuration in $\mathcal{D}_\mathrm{Test}^{\mathrm{top}\,75\%}$ is less than $10^{-7}$ for $N=16$. We believe this observation explains the more subdued double descent observed in \cref{fig:N16}(a) and (b). 

Since the Hilbert space is exponentially larger, any realistic NQS experiment would involve a number of training samples that corresponds to a much smaller percentage of the Hilbert space. As such, 
we perform a second set of experiments where we train our NQS on $\mathcal{D}_\mathrm{Train}^{\mathrm{top}\,25\%}$, which consists of only the $25\%$ of configurations with the largest ground-state wavefunction amplitudes. Using a smaller training set also allows us to test our hypothesis about the effect of having only very small probabilities in the test set. In this case, more of the Hilbert space is included in the test set, so the average probability of a configuration is an order of magnitude larger. We note that we train these NQS for 30,000 epochs. \Cref{fig:N16}(c) shows the training and test loss achieved by these models. In this case, the test loss and the infidelities shown in (d) exhibit much more pronounced double descent behavior. Again, we identify our best estimate of the interpolation threshold.

For both sets of experiments we observe that the interpolation threshold, which marks the transition to the overparameterized regime where networks learn to generalize well, is located for $N_\mathrm{params} > 2^N$. Therefore, these experiments support our conclusions in the main text: In our setting, where our NQS are trained with a supervised learning procedure, networks with a desirable number of parameters, i.e., $N_\mathrm{params}<2^N$, operate in the underparameterized regime.

\section{The Hellinger distance}
\label{app:Hellinger}

Formally, a probability space is defined as the following triplet: a sample space, an event space (a subset of the sample space), and a probability function that assigns a probability to the events in the event space.
In this work, the sample space is the full Hilbert space, the event space is the set of spin configurations in the training set, and the probability function is given by a wavefunction according to the Born rule. Together, these elements define the probability space,
\begin{align*}
    \Big(\left\{\vec{\sigma}\right\}, \left\{\vec{\sigma}\in\mathcal{D}_\mathrm{Train}\right\}, p = \vert\langle\Psi\vert\Psi\rangle\vert^2\Big).
\end{align*}
One distribution in this probability space is the Born distribution corresponding to the exact ground-state wavefunction: $p_\Omega = \big\vert\langle\Omega\vert\Omega\rangle\big\vert^2$. Other distributions in the space include the Born distributions corresponding to the wavefunctions learned by our NQS: $p_\theta = \big\vert\langle\Psi_\theta\vert\Psi_\theta\rangle\big\vert^2$.

The Hellinger distance between two distributions $P = \vert\langle\Psi_P\vert\Psi_P\rangle\vert^2$ and $Q = \vert\langle\Psi_Q\vert\Psi_Q\rangle\vert^2$ in our probability space is defined as
\begin{align*}
    \mathcal{H}(P,Q)  &= \frac{1}{\sqrt{2}}\sqrt{
    \sum_{\left\{\vec{\sigma}\right\}} 
    \Bigg(\sqrt{P(\vec{\sigma})} - \sqrt{Q(\vec{\sigma})}\Bigg)^2
    }.
\end{align*}
Note that the sum is over the full sample space, which is the full Hilbert space $\left\{\vec{\sigma}\right\}$. 
If $\vert\Psi_P\rangle$ and $\vert\Psi_Q\rangle$ have only real and non-negative amplitudes, then $\Psi_P(\vec{\sigma})\equiv\sqrt{P(\vec{\sigma})} $ and $\Psi_Q(\vec{\sigma}) \equiv\sqrt{Q(\vec{\sigma})}$, and
the Hellinger distance can be equivalently defined as 
\begin{align*}
    \mathcal{H}(P,Q)  & = \frac{1}{\sqrt{2}}\sqrt{
    \sum_{\left\{\vec{\sigma}\right\}}
    \big(\Psi_P(\vec{\sigma}) - \Psi_Q(\vec{\sigma})\big)^2
    }.
\end{align*}

In this work, we train NQS to represent a target quantum state $\vert\Omega\rangle$, which is the ground state of the Hamiltonian defined in Eq. 2 in the main text. This ground state is known to have only real and non-negative amplitudes. As such, we restrict our NQS to also have only real and non-negative amplitudes. Therefore, if our trained NQS $\vert\Psi_\theta\rangle$ accurately represents $\vert\Omega\rangle$, then the distance between $p_\theta$ and $p_\Omega$, or equivalently,
\begin{align*}
    \mathcal{H}(\Psi_\theta,\Omega) = \frac{1}{\sqrt{2}}\sqrt{
    \sum_{\left\{\vec{\sigma}\right\}}
    \big(\Psi_\theta(\vec{\sigma}) - \Omega(\vec{\sigma})\big)^2
    }, 
\end{align*}
will be small. Indeed, $\mathcal{H}(\Psi_\theta,\Omega)$ is proportional to the $L^2$ norm between the learned wavefunction amplitudes and the true ground-state wavefunction amplitudes.  Viewed through a machine learning lens, this distance takes the form of a rescaled squared error. Based on this observation, we designed the loss function defined by Eq. 1 in the main text, where we replace the sum over the entire Hilbert space $\left\{\vec{\sigma}\right\}$ with a sum over the spin configurations in the training set $\vec{\sigma}\in\mathcal{D}_\mathrm{Train}$. Furthermore, we replace the normalized amplitudes of the learned wavefunction $\Psi_\theta(\vec{\sigma})$ with the unnormalized ones $\psi_\theta(\vec{\sigma})$. As a consequence, our loss function $\mathcal{L}_\mathcal{H}$ is only bounded from below, whereas the $\mathcal{H}(\Psi_\theta,\Omega)$ is also bounded from above by $1/\sqrt{2}$.

Another popular quantity to measure differences between probability distributions is the Kullback-Leibler (KL) divergence~\cite{kullback1951information-68f}. However, the KL divergence, in contrast to the Hellinger distance, is not a proper metric (it is not symmetric and does not satisfy the triangle inequality). 
We note that the metric corresponding to the general class of quantum states, which are not necessarily real or non-negative, is the Fubini-Study metric. This metric is central to stochastic reconfiguration, an optimization technique tailored to NQS. We emphasize that the Hellinger distance is a suitable choice for our task because we are easily able to transform the metric into a squared error loss function, similar to those commonly used in classical machine learning practice. For the Fubini-Study metric, on the other hand, such a transformation is less obvious.

The mean-squared error (MSE) loss function is one of the most common loss functions in the classical machine learning literature. For that reason, we attempted to perform our experiments using the MSE loss. However, we observed that minimizing the MSE loss led to instabilities. This is likely due to the lack of a square root in the MSE formulation (i.e., the square roots of the individual probabilities and the square root of the summed errors). We found that the Hellinger distance better handles wavefunction amplitudes that span several orders of magnitude (see, e.g., the $y$-axis in \cref{fig:datasets}).

\section{Details of the optimization}
\label{app:optimization}

For all of the experiments shown in the main text, we train our NQS by minimizing the loss function inspired by the Hellinger distance, defined in Eq. 1 in the main text. We compute the loss and gradients of the loss on small batches of the training configurations for each training iteration. We observe that batching introduces beneficial noise into the gradients and helps our networks converge faster. For $N=12$, we fix the batch size to $2^6$, unless stated otherwise. We emphasize, however, that in a single training epoch, which consists of multiple gradient steps, the neural network sees every configuration in the training set. We train all NQS for 15,000 epochs, except for NQS trained on $\mathcal{D}_\mathrm{Train}^{\mathrm{IS}\,75\%}$ (see \cref{app:IS_data}) and NQS trained on $\mathcal{D}_\mathrm{Train}^{\mathrm{top}\,25\%}$ for $N=16$ (see \cref{app:N16}). 

During the training, we exponentially decay the learning rate such that, at a given training epoch $t$, the learning rate is given by a schedule that depends on a maximum learning rate value $\lambda_\mathrm{max} $, a decay rate $r$, and a number of transition steps $N_\mathrm{trans.}$. The schedule is defined as
    \begin{align*}
        \lambda(t) = \lambda_\mathrm{max} \times r ^{t N_\mathrm{trans}^{-1}}.
    \end{align*}
We set $\lambda_\mathrm{max}  = 0.001$, $r = 0.99$, and $N_\mathrm{trans.}=1000$.

\begin{figure}
    \centering
    \includegraphics[width=\linewidth]{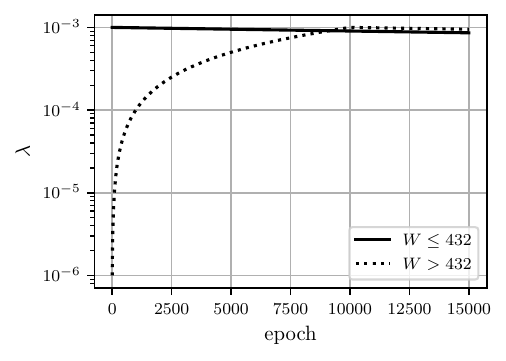}
    \caption{During the training of our NQS, we adjust the value of the learning rate according to the two schedules shown above. For networks with $W \leq 432$ (solid line), we exponentially decay the learning rate from an initial value of $0.001$. For wider networks (dotted line), we first linearly increase the learning rate from $10^{-6}$ and then decrease it exponentially.}
    \label{fig:learning_rates}
\end{figure}

For very wide networks, training is very sensitive to the learning rate schedule. For networks with $W>432$, we found that the \emph{training loss} would sometimes increase as a function of network size or that some random initializations would cause the optimization to get stuck after the first training epoch. As such, we adjusted the learning rate schedule for networks with $W> 432$ so that the learning rate linearly ``warms up'' from an initial value $\lambda_0 = 10^{-6}$ to a maximum value $\lambda_\mathrm{max} = 0.001$ during the first $10\times10^3$ training epochs. For the remainder of the training, the learning rate follows the exponential decay defined above. The two learning rate schedules are displayed in \cref{fig:learning_rates}. Notably, the very wide networks with $W>432$ behave differently for different training datasets. In particular, the learning rate schedule we introduced for the networks with $W>432$ only stabilized the training of those networks for the experiments summarized in Figs. 1 and 6 in the main text. For our other experiments, such as those summarized in Figs. 3 and 5 in the main text and \cref{fig:para_ferro_dd}, we focus on networks with $W\leq432$. The double descent behavior that these experiments support can be clearly seen without considering wider networks.

\section{Details of the training data}
\label{app:data}
\begin{figure*}
\includegraphics[width=\textwidth]{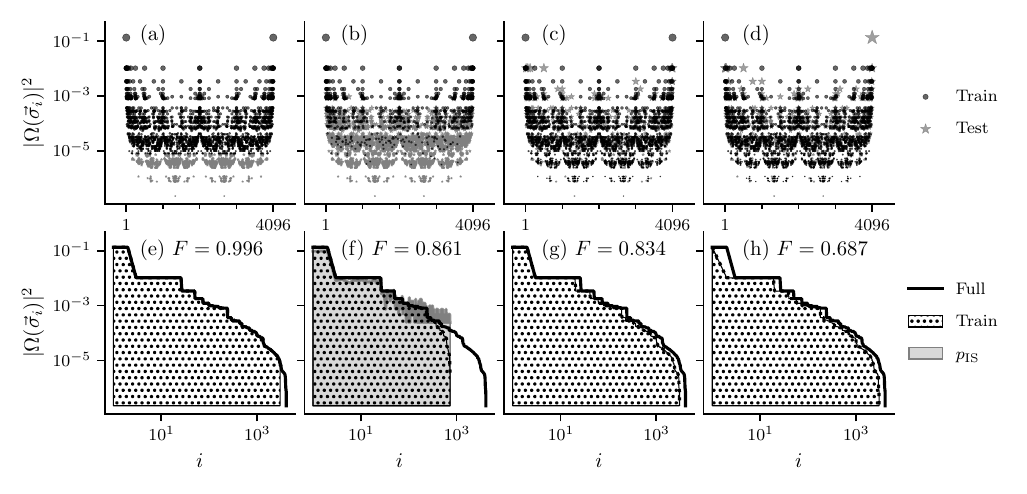}
    \caption{Panels (a)-(d) show the different ways we split the full set of spin configurations $\left\{\vec{\sigma}\right\}$ into training and test sets, and panels (e)-(h) show the  probability density of $p_\Omega$ contained in each training dataset. For all choices of the training dataset, $\mathcal{D}_\mathrm{Train}$, the test dataset, $\mathcal{D}_\mathrm{Test}$, contains the spin configurations not in the training dataset: $\mathcal{D}_\mathrm{Test} = \{\vec{\sigma}\}\,\backslash\,\mathcal{D}_\mathrm{Train}$. (a), (e) Data split according to the true Born distribution. The training dataset, $\mathcal{D}_\mathrm{Train}^{\mathrm{top}\,75\%}$, contains the $N_s= 75\%\times2^N$ configurations with the largest amplitudes in the true ground-state wavefunction $\vert\Omega\rangle$. (b), (f) Training data importance-sampled from the true Born distribution (i.e. sampled with replacement). While the training dataset, $\mathcal{D}_\mathrm{Train}^{\mathrm{IS}\,75\%}$, again contains the $N_s= 75\%\times2^N$ configurations, only a small fraction of those configurations are unique. (c), (g) Training data sampled (without replacement) from a uniform distribution over all configurations. For this seed, both high-probability configurations are in the training dataset, $\mathcal{D}_\mathrm{Train}^{\mathrm{unif.}\,75\%}$. (d), (h) Training data sampled (without replacement) from a uniform distribution over all configurations. Here, only one of the high-probability configurations is in $\mathcal{D}_\mathrm{Train}^{\mathrm{unif.}\,75\%}$.}
    \label{fig:datasets}
\end{figure*}

Access to a target ground state $\vert\Omega\rangle$ is equivalent to knowing the wavefunction amplitudes for all $2^N$ $z$-basis states: $\Omega(\vec{\sigma}) = \langle\vec{\sigma}\vert\Omega\rangle \,\,\forall\,\,\vec{\sigma}\in\left\{\vec{\sigma}\right\}$. We can split this set of spin configurations and their corresponding amplitudes into training and test sets according many different protocols. Here we describe in more detail all of the dataset splittings considered in this manuscript. \Cref{fig:datasets} provides two visualizations for each type of dataset. The top row shows how individual spin configurations are categorized into a training set, $\mathcal{D}_\mathrm{Train}$, and a test set, $\mathcal{D}_\mathrm{Test}$. The bottom row  of \cref{fig:datasets} displays the portion of the probability density function (PDF) corresponding to the true Born distribution $p_\Omega$ that is contained in $\mathcal{D}_\mathrm{Train}$. 

We first consider the ``best'' case for the neural network, where the training data contains the most information about the true Born distribution $p_\Omega$, and therefore the target wavefunction $\vert\Omega\rangle$. 
To create such a training set, we sort the configurations and their amplitudes according to $p_\Omega$. We then take the 75$\%$ of configurations with the largest probabilities (and thus the largest amplitudes) as the training set, which we refer to as $\mathcal{D}_\mathrm{Train}^{\mathrm{top}\,75\%}$. The remaining 25$\%$ of configurations have the smallest probabilities, and make up the test set, which we respectively refer to as $\mathcal{D}_\mathrm{Test}^{\mathrm{top}\,75\%}$.
We display this data splitting in \cref{fig:datasets} (a) and the probability density of $p_\Omega$ contained in $\mathcal{D}_\mathrm{Train}^{\mathrm{top}\,75\%}$ in panel (e). 
We used this dataset to obtain the results presented in Figs. 1, 2, and 4 in the main text. Only the purple lines in Fig. 4 correspond to NQS trained on this type of dataset. For experiments in \cref{app:less_data}, we use this data splitting protocol but with different ratios between the training and test sets, namely 25:75 ($\mathcal{D}_\mathrm{Train}^{\mathrm{top}\,25\%}$ and $\mathcal{D}_\mathrm{Test}^{\mathrm{top}\,25\%}$) and 50:50 ($\mathcal{D}_\mathrm{Train}^{\mathrm{top}\,50\%}$ and $\mathcal{D}_\mathrm{Test}^{\mathrm{top}\,50\%}$).

While the above data splitting protocol produces training sets with maximal information about the target wavefunction and its corresponding Born distribution, this situation is rarely encountered when training NQS. 
During the variational training of NQS, the configurations used for training are generated by \emph{importance sampling} from the NQS distribution itself.
In order to more closely resemble this setting, we importance sample from the true Born distribution $p_\Omega$. In other words, we ``sample with replacement'' from $p_\Omega$. For each random seed, the generated training sets are different, while still capturing the majority of the high-probability configurations. We sample $N_\mathrm{samples} = 75\% \times 2^N$ configurations, but since we sample with replacement, the number of unique configurations is much smaller than $N_\mathrm{samples}$. 
Training datasets generated via importance sampling are referred to $\mathcal{D}_\mathrm{Train}^{\mathrm{IS}\,75\%}$ and the test datasets contain all configurations that are not in the training set. In other words, $\mathcal{D}_\mathrm{Test}^{\mathrm{IS}\,75\%} = \{\vec{\sigma}\}\,\backslash\,\mathcal{D}_\mathrm{Train}^{\mathrm{IS}\,75\%}$. 
We present an example of one such $\mathcal{D}_\mathrm{Train}^{\mathrm{IS}\,75\%}$ in \cref{fig:datasets}(b) and the probability density of $p_\Omega$ contained in $\mathcal{D}_\mathrm{Train}^{\mathrm{IS}\,75\%}$ in panel (f). 
The frequency with which a given configuration is sampled, normalized by the total number of samples $N_\mathrm{samples}$, provides an estimate of the true probability of that sample. Based on this, we define 
\begin{align*}
    p_\mathrm{IS}(\vec{\sigma}^*) = \frac{1}{N_\mathrm{samples}} \sum_{\left\{\vec{\sigma} \in \mathcal{D}_\mathrm{Train}\right\}} \delta_{\vec{\sigma},\vec{\sigma}^*},
\end{align*}
which, in our case, is the importance-sampled approximation of $p_\Omega(\vec{\sigma}^*)$.
\Cref{fig:datasets} (f) shows that this empirical distribution closely matches the portion of the true Born distribution PDF that is contained in $\mathcal{D}_\mathrm{Train}^{\mathrm{IS}\,75\%}$.
We used the datasets generated via importance sampling to obtain the results presented in \cref{app:IS_data}.

To contrast with the data splitting protocols described thus far, which make use of the true Born distribution $p_\Omega$, we also generate training and test sets by randomly splitting spin configurations and their amplitudes between the training and test sets. 
This protocol is synonymous with sampling configurations (without replacement) from a uniform distribution, and thus, it is agnostic to the true ground-state wavefunction. 
In this work, we consider training datasets, $\mathcal{D}_\mathrm{Train}^{\mathrm{unif.}\,75\%}$, with $75\%$ of all spin configurations sampled from the uniform distribution. This type of training dataset resembles samples drawn from a randomly initialized NQS. The corresponding test datasets contain the spin configurations not in the training dataset, i.e. $\mathcal{D}_\mathrm{Test}^{\mathrm{unif.}\,75\%}  = \{\vec{\sigma}\}\,\backslash\,\mathcal{D}_\mathrm{Train}^{\mathrm{unif.}\,75\%}$.
Datasets generated with different seeds can have a very different quality, which manifests in the training process, as described in the main text. In particular, the quality of a given generated training set hinges on how many high-probability configurations the training set contains. For the critical TFIM ($h=1$), there are two spin configurations with particularly high probabilities which dominate the Born distribution. Each dataset generated for our experiments either contained both of these configurations (see an example in panels (c), (g) in \cref{fig:datasets}) or only one of them  (an example in panels (d), (h) in \cref{fig:datasets}). We used these datasets to obtain the results presented in Figs. 3 and 4 in the main text. The orange lines in Fig. 4 correspond to NQS trained on this type of dataset.

In order to more concretely compare the datasets produced with the described protocols, we also consider the fraction of the PDF corresponding to $p_\Omega$ present in each training dataset. This fraction is defined as
\begin{align*}
    F = \frac{\sum_{\left\{\vec{\sigma}^\prime \in \mathcal{D}_\mathrm{Train}\right\}} \vert\Omega(\vec{\sigma}^\prime)\vert^2}
    {\sum_{\left\{\vec{\sigma}\right\}} \vert\Omega(\vec{\sigma})\vert^2}. 
\end{align*}
As mentioned, the first data splitting we consider produces datasets with the most information about $p_\Omega$, with a value of $F=0.996$. Even though datasets produced with importance sampling, or sampling with replacement, still contain the configurations with the highest probabilities, there is significantly less information about the target distribution with a value of $F=0.861$ for the seed shown in \cref{fig:datasets}. This is a direct consequence of the fact that the number of unique configurations in the training set is significantly less than the total size of the training set. Interestingly, the value of $F$ for the randomly generated training set shown in \cref{fig:datasets} (c), (g), $F=0.834$, is not much smaller than the dataset generated via importance sampling. Notably, that dataset contains both high-probability configurations, as seen in \cref{fig:datasets} (c).  If only one of the high-probability configurations is in the training set, as is the case for the randomly generated dataset shown in \cref{fig:datasets} (d), (h), then the value of $F$ drops to $F=0.687$. 

\section{Details of the neural network architecture}
\label{app:architecture}

\begin{figure}
    \centering
    \includegraphics[width=\linewidth]{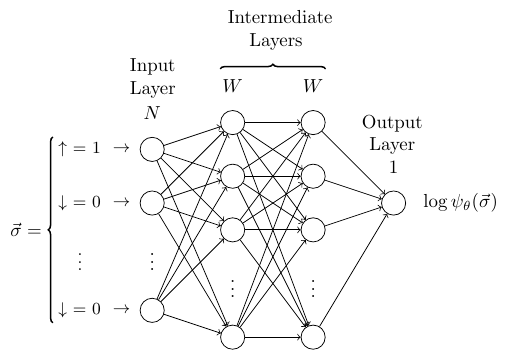}
    \vspace{0.3em}
    \caption{We employ a three-layer feed-forward neural network architecture with variable width $W$. This neural network has $N$ input nodes, corresponding to the number of spins in the physical system. The input to these nodes is a spin configuration $\vec{\sigma} = (\sigma_1,\sigma_2,\dots,\sigma_N)$, which is a $\sigma^z$-basis state, and the network outputs the logarithm of the wavefunction amplitude associated with that spin configuration $\mathrm{log}\,\psi_\theta(\vec{\sigma})$.}
    \label{fig:architecture}
\end{figure}
\begin{figure}
    \centering
    \includegraphics[width=\linewidth]{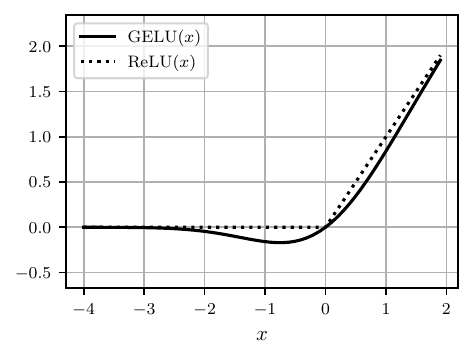}
    \caption{We define our neural network with a Gaussian Error Linear Unit (GELU) activation function. This is a smoother alternative to the rectified linear unit (ReLU) activation function, which is shown for reference. }
    \label{fig:activation}
\end{figure}

For all of the experiments presented in this work, we employ a three-layer feed-forward neural network as our NQS architecture, as shown in \cref{fig:architecture}. Each network has $N$ input nodes, equal to the number of spins in the system, and a single output node. The width of the intermediate layers $W$ is adjusted to control the total number of trainable parameters in the network $N_\mathrm{params}$. Because we fix the depth of the network to three, $N_\mathrm{params}$ depends only on the number of spins in the physical system $N$ and the width of the intermediate layers $W$. 

\begin{figure*}
    \centering
    \includegraphics[width=0.9\textwidth]{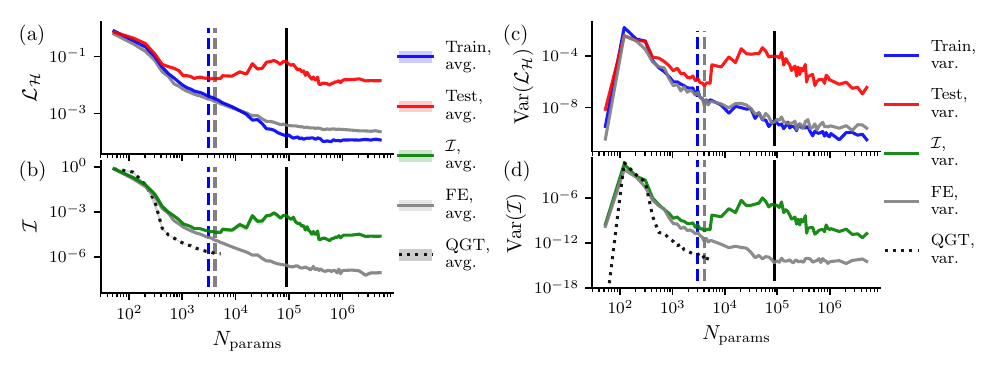}
    \caption{
    The colored lines show the average behavior of the training loss, test loss, and infidelity from Fig. 1 (b)-(c) in the main text, where our NQS were trained on $\mathcal{D}_\mathrm{Train}^{\mathrm{top}\,75\%}$. The grey line shows the average behavior of our NQS trained on the complete set of spin configurations. The black dotted line in (b) shows the infidelity achieved with our NQS when the loss function is the infidelity and the optimization makes use of the quantum geometric tensor. The variances of (c) the training and test losses and (d) the infidelities displayed in (a) and (b), respectively. The black vertical line represents our estimate of the interpolation threshold from Fig. 1. The gray dashed line (blue dashed line) indicates where the number of network parameters equals the size of the Hilbert space, $N_\mathrm{params}=2^N$ (the number of training configurations in $\mathcal{D}_\mathrm{Train}^{\mathrm{top}\,75\%}$, $N_\mathrm{params}=75\% \times 2^N$).}
    \label{fig:FE}
\end{figure*}

Each intermediate layer of the network consists of three steps: the computation of the \emph{pre-activation}, an affine transformation of the inputs to the layer, then a layer normalization of the pre-activation, and finally the application of a non-linear activation function. The affine transformation is a linear transformation involving the trainable weights of the layer. The layer normalization helps mitigate training instabilities for very wide networks. We employ a Gaussian Error Linear Unit (GELU) activation function for all intermediate layers. The GELU activation function, shown in \cref{fig:activation}, is a smoother alternative to the commonly used Rectified Linear Unit (ReLU) activation function. For networks with ReLU-like activation functions, such as GELU, all trainable weights $\theta$ are initialized according to the He initialization~\cite{he2015delving-b9b}. This initialization strategy helps prevent vanishing and exploding gradients, especially for very deep or wide networks. Importantly, there is no layer normalization or activation function in the output layer. The final output of our neural network architecture is a single number, which we interpret as the logarithm of the unnormalized wavefunction amplitude of the input spin configuration.

\section{Expressiveness of the NQS}
\label{app:expressiveness}

While the main text focuses on generalization and trainability, here we examine the expressivity of our NQS as a function of the number of parameters. First, we train the NQS by minimizing the loss function evaluated on the complete set of spin configurations, without a partitioned test set. This optimization is carried out using the same details described in the previous section titled ``Details of the optimization''. In particular, gradients are computed in batches of size $2^6$ and parameters are updated according to those gradients using the Adam optimizer~\cite{kingma_adam_2017}. As shown in \cref{fig:FE} (grey lines), for intermediate and large network widths, the learned wavefunction is a better approximation of the true ground state, as measured by infidelity, despite a slightly higher training loss compared to the setup in Fig. 1 (b)-(c) in the main text (colored lines). We attribute this increase in training loss to the larger number of training configurations, as the loss is computed as a sum over the spin configurations, not an average. Importantly, the quality of the approximation does not exhibit any double descent behavior, indicating that the peaks in the test loss in the main text come from factors other than the expressivity of the ansatz. 

We also train our NQS by minimizing the infidelity with respect to the target wavefunction. In this case, we optimize the parameters of the neural network with natural gradient descent, following the procedure outlined in Ref.~\cite{dash2025efficiency-013}. 
The infidelities achieved using this optimization strategy are shown in  \cref{fig:FE}(b) (dotted line). For network sizes where it is possible, this optimization scheme leads to the lowest infidelities, which improve quickly with the network size. 
However, for networks with $N_\mathrm{params}>2^N$, the construction and inversion of the quantum geometric tensor (QGT), a central step of natural gradient descent, becomes a memory bottleneck.
Note that we do not batch our training configurations for this type of optimization, since it is imperative to normalize the neural network amplitudes when computing the infidelity and the QGT.  We use a constant learning rate of $\lambda = 0.01$ and we apply a diagonal shift to the QGT, which stabilizes the inversion. We exponentially decay this diagonal shift from an initial value of $\delta_0 = 0.01$ using a decay rate of 0.99 and 100 transition steps.

\section{Correlation functions}
\label{app:correlations}

\begin{figure*}
    \centering
    \includegraphics[width=0.8\textwidth]{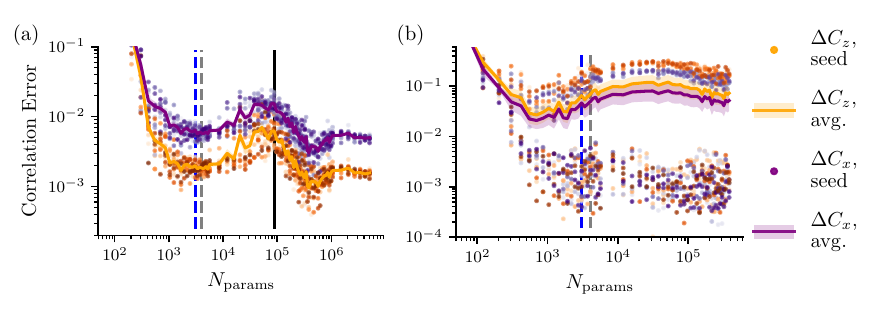}
    \caption{The correlation error in $x$ and $z$ for (a) the NQS trained on $\mathcal{D}_\mathrm{Train}^{\mathrm{top}\,75\%}$ (corresponding to Fig. 1 (b)-(c) in the main text) (b) the NQS trained on $\mathcal{D}_\mathrm{Train}^{\mathrm{unif.}\,75\%}$ (corresponding to Fig. 3 in the main text). Markers represent individual trained networks, and the solid lines represent the averages over ten random initializations. The vertical lines follow the same convention as in \cref{fig:FE}.}
    \label{fig:corrs}
\end{figure*}

In addition to the double descent behavior in the test loss and infidelity in Fig. 1 (b)-(c) in the main text, similar features appear in other physically meaningful quantities. In particular, this behavior appears for spin–spin correlations in the $z$ and $x$ directions between sites separated by a distance $r$,
\begin{align*}
    C_z(r) =\frac{1}{L} \sum_{i}\langle S_i^zS_{i+r}^z\rangle, \\ C_x(r) =\frac{1}{L} \sum_{i}\langle S_i^xS_{i+r}^x\rangle,
\end{align*}
To summarize the network's ability to capture the correlations of the target ground-state wavefunction $\vert\Omega\rangle$, we consider the $z$ and $x$ correlation error, defined as the total deviation from the exact values for $r = 1-5$, 
\begin{align*}
    \Delta C_z= \sum_{r=1}^5 |C_z(r) - C_z^{\mathrm{exact}}(r)|, \\ \Delta C_x= \sum_{r=1}^5 |C_x(r) - C_x^{\mathrm{exact}}(r)|,
    \label{eq:corr_err}
\end{align*}
where $C_z^{\mathrm{exact}}(r)$ and $C_x^{\mathrm{exact}}(r)$ are computed using  the exact ground state $\vert\Omega\rangle$. While the infidelity between the learned ground state and the target ground state can be used to bound the error on any other observable estimated with the learned ground state~\cite{Becca_book_2017,beach_making_2019}, the behavior of the correlations in different bases sheds new light on how our trained NQS overfit to the training data.

\Cref{fig:corrs}(a) displays the correlation errors for the NQS trained on the $75\%$ of configurations with the largest exact ground-state wavefunction amplitudes. The correlations are estimated using the trained NQS which produced the results presented in Fig. 1 (b)-(c) in the main text. In this case, the $z$-correlation error is consistently lower than that of the $x$-correlation. This is likely related to the fact that the network is trained on the highest-probability configurations, which are $z$-basis states and contribute the most to the $z$-basis correlations.
\Cref{fig:corrs}(b) shows the correlation errors for the NQS trained on the training sets generated uniformly at random. The correlations are estimated using the trained NQS which produced the results presented in Fig. 3 in the main text.  
As with the test loss, the correlation errors for these NQS depend on the specific training data. Seeds where the dataset includes only a single high-probability configuration have higher correlation errors. In these cases, the trained NQS capture the $x$ correlations slightly more accurately compared to the $z$ correlations. Fig. 4 (a) in the main text shows that, for these seeds, the trained NQS severely underestimate amplitudes of the test set, leading to an underestimated normalization constant $\mathcal{N}$. In particular, the network learns an inaccurate amplitude of the high-probability test configuration, which more severely affects the $z$ correlation error as compared to the $x$ correlation error. These correlation errors highlight how the choice of basis for training NQS may bias the errors on basis-dependent observables.

\section{Further investigation into the parity error}
\label{app:parity}

We introduced the parity error $\epsilon_\mathrm{parity}$ to measure how well-trained NQS respect the ground state's parity symmetry. We repeat the definition of $\epsilon_\mathrm{parity}$ here for convenience,
\begin{equation}\label{eq:parity_error_app}
    \epsilon_{\mathrm{parity}}= \frac{1}{|\mathcal{D}_{\mathrm{parity}}|}\sum_{\mathcal{D}_{\mathrm{parity}}} \left\vert1-  \frac{\psi_{\theta}(\mathcal{P}\vec{\sigma})}{\psi_{\theta}(\vec{\sigma})}\right\vert.
\end{equation}
Recall that $\mathcal{P}$ is a parity operator that flips all of the spins in a configurations $\vec{\sigma}$.
In the main text, we computed the parity error over all parity-symmetric pairs in the test dataset, i.e., $\mathcal{D}_{\mathrm{parity}} = \left\{ \vec{\sigma} \;\middle|\; \vec{\sigma}, \mathcal{P}\vec{\sigma} \in \mathcal{D}_{\mathrm{Test}} \right\}
$. \Cref{fig:parity_norm} shows those results. Curiously, the NQS trained on $\mathcal{D}_{\rm Train}^{\mathrm{unif.}\,75\%}$ did not show double descent behavior in $\epsilon_\mathrm{parity}$, meaning the parity learning improves with increased network size, even when NQS overfit to the training data. 
To better understand our observation, we calculate $\epsilon_\mathrm{parity}$ using new definitions of $\mathcal{D}_\mathrm{parity}$. We focus on the same networks, namely NQS trained on $\mathcal{D}_{\rm Train}^{\mathrm{unif.}\,75\%}$ datasets containing only one of the two high-probability configurations.

\begin{figure}
    \centering
    \includegraphics[width=\linewidth]{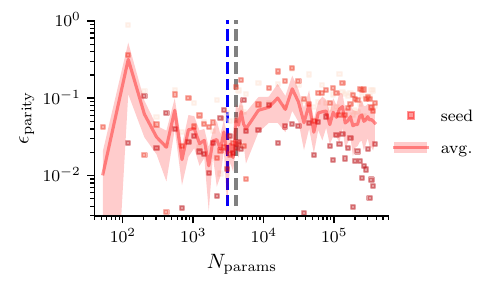}
    \caption{The parity error $\epsilon_\mathrm{parity}$ of NQS trained on $\mathcal{D}_\mathrm{Train}^{\mathrm{unif.}\,75\%}$, specifically the datasets containing only one of the two high-probability configurations. Here, we examine the parity error (\cref{eq:parity_error}) only for parity-symmetric pairs in the test dataset with the largest ground-state amplitudes (\cref{eq:second_row_test}). 
    Markers represent individual trained networks, and the solid lines represent the averages over random initializations. The vertical lines follow the same convention as in \cref{fig:FE}.}
\label{fig:second_row_test}
\end{figure}

As a first step, we examine the parity error for \emph{specific} parity-symmetric pairs in the test dataset.
For example, \Cref{fig:second_row_test} shows the parity error measured over parity-symmetric pairs in the test dataset with the largest ground-state amplitudes: 
\begin{align}
    \label{eq:second_row_test}
    \mathcal{D}_\mathrm{parity} = \big\{ \vec{\sigma} \,\vert\, &\nonumber\vec{\sigma}, \mathcal{P}\vec{\sigma} \in \mathcal{D}_{\mathrm{Test}}, \\
    &\Omega(\vec{\sigma}) = \Omega(\mathcal{P}\vec{\sigma})\geq \Omega(\vec{\sigma}^\prime) \,\forall\, \vec{\sigma}^\prime \in \mathcal{D}_\mathrm{Test}
    \big\}.
\end{align} 
Note that this is a subset of the original definition of $\mathcal{D}_\mathrm{parity}$. When restricting our parity error to this specific set of configurations, we see a subtle signature of double descent: $\epsilon_\mathrm{parity}$ increases for the intermediate-sized NQS that exhibit overfitting in \cref{fig:random_DD} and then slightly improves again for larger networks. This observation suggests that while the parity error, on average (\cref{fig:parity_norm}), improves with increased network size, the parity error for specific, high-probability pairs suffers when the NQS overfit.

\begin{figure*}
    \centering
    \includegraphics[width=0.66\textwidth]{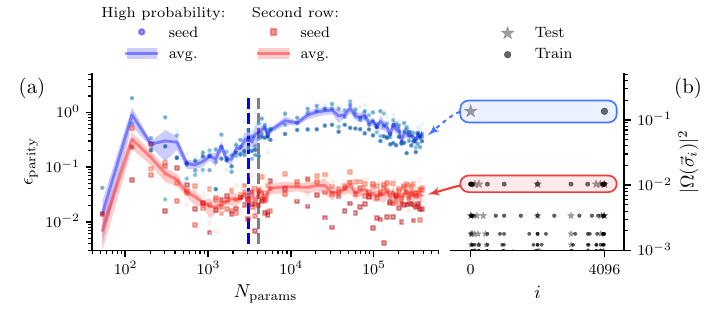}
    \caption{(a) The parity error $\epsilon_\mathrm{parity}$ of NQS trained on $\mathcal{D}_\mathrm{Train}^{\mathrm{unif.}\,75\%}$, specifically the datasets containing only one of the two high-probability configurations. Here, we examine the parity error (\cref{eq:parity_error_app}) for specific parity-symmetric pairs that are split between the training and test datasets (\cref{eq:train_test}). First we consider the parity error for the high-probability configurations, which are connected by $\mathcal{P}$ (shown in blue). We emphasize that for the seeds shown here, one high-probability configuration is in $\mathcal{D}_\mathrm{Train}^{\mathrm{unif.}\,75\%}$ and the other is in $\mathcal{D}_\mathrm{Test}^{\mathrm{unif.}\,75\%}$. We also consider the parity error between pairs that satisfy the criterion in \cref{eq:train_test} and have the second largest ground-state amplitude (shown in red). 
    Markers represent individual trained networks, and the solid lines represent the averages over random initializations. The vertical lines follow the same convention as in \cref{fig:FE}.
    Panel (b) shows the largest squared wavefunction amplitudes with dots and stars indicating the training and test configurations, respectively, for a single, exemplary instance of $\mathcal{D}_\mathrm{Train}^{\mathrm{unif.}\,75\%}$ and $\mathcal{D}_\mathrm{Test}^{\mathrm{unif.}\,75\%}$. The blue and red shaded regions highlight the high-probability configurations and the second row of the Born distribution, which were used to compute the parity errors shown in panel (a).
    }
    \label{fig:train_test_specific}
\end{figure*}

\begin{figure}
    \centering
    \includegraphics[width=\linewidth]{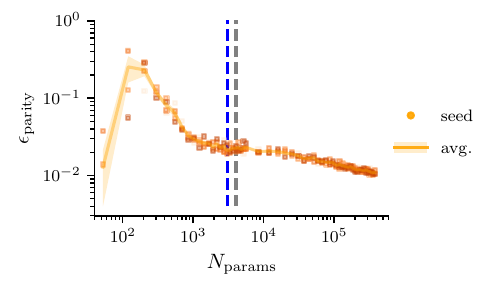}
    \caption{The parity error $\epsilon_\mathrm{parity}$ of NQS trained on $\mathcal{D}_\mathrm{Train}^{\mathrm{unif.}\,75\%}$, specifically the datasets containing only one of the two high-probability configurations. Here, we examine the parity error (\cref{eq:parity_error_app}) averaged over
    parity-symmetric pairs that are split between the training and test datasets (\cref{eq:train_test}).
    Markers represent individual trained networks, and the solid lines represent the averages over random initializations. The vertical lines follow the same convention as in \cref{fig:FE}.}
    \label{fig:test_train_all}
\end{figure}

For the datasets considered in this appendix, the parity-symmetric pairs in the test dataset with the largest ground-state amplitudes have the second largest amplitudes overall. In other words, these pairs belong to the ``second row'' of the Born distribution corresponding to the true ground-state wavefunction (highlighted in red in \cref{fig:train_test_specific} (b)). This is because one of the configurations in the high-probability parity pair is always in the training dataset. Even though there are many parity-symmetric pairs in this second row, only one or two of those pairs satisfy the condition that both $\vec{\sigma}$ and $\mathcal{P}\vec{\sigma}$ are contained in the test dataset, for the datsets considered here. As such, the results in \cref{fig:second_row_test} are noisy, making the double descent behavior difficult to characterize.

We now consider the parity error measured for parity-symmetric pairs that are split between the training and test datasets:
\begin{align}
    \label{eq:train_test}
    \mathcal{D}_\mathrm{parity} = \big\{ \vec{\sigma} \,\vert\, \vec{\sigma} \in \mathcal{D}_{\mathrm{Train}}, \mathcal{P}\vec{\sigma} \in \mathcal{D}_{\mathrm{Test}} \big\}
\end{align}
\Cref{fig:test_train_all} shows the parity error averaged over all parity-symmetric pairs that are split between the test and train datasets. 
We expected that the parity error evaluated over these pairs would reflect the overfitting seen in \cref{fig:random_DD}, since one configuration in every pair is in the training dataset to which the NQS overfit. 
However, we again see that $\epsilon_\mathrm{parity}$ monotonically decreases with increasing network size, as was the case in \cref{fig:parity_norm} (b).

Considering pairs split between the training and test datasets also allows us to further test the hypothesis that the parity error between symmetric configurations $\vec{\sigma}$ and $\mathcal{P}\vec{\sigma}$ depends on the magnitude of their amplitudes $\vert\Omega(\vec{\sigma})\vert =\vert\Omega(\mathcal{P}\vec{\sigma})\vert$.
The training datasets considered here contain one of the high-probability configurations, meaning we can examine the parity error for the high-probability pair (highlighted in blue in \cref{fig:train_test_specific} (b)). Furthermore, there are more parity-symmetric pairs in the second row of the Born distribution (highlighted in red in \cref{fig:train_test_specific} (b)) that satisfy the criterion in \cref{eq:train_test}. \Cref{fig:train_test_specific} (a) displays $\epsilon_\mathrm{parity}$ for these specific parity-symmetric pairs. Both parity errors show clear double descent behavior, with $\epsilon_\mathrm{parity}$ increasing for the intermediate-sized NQS that overfit to the training data and then improving again for larger networks. 
Notably, the double descent behavior is more pronounced for $\epsilon_\mathrm{parity}$ evaluated on the high-probability pair. Therefore, we conclude that the parity error between symmetric configurations depends on the magnitude of their ground-state amplitudes.

Our analysis shows that the ground state's parity symmetry is continuously learned for increasing network size, but that the learned symmetry is sacrificed or ignored for configurations with large ground-state amplitudes when the NQS overfit. 
In other words, our NQS tend to overfit more severely to configurations with large ground-state amplitudes. This could be explained by the fact that our loss function, defined in \cref{eq:loss}, is a function of squared errors, which are \emph{absolute} errors. Large ground-state amplitudes yield more signal during optimization, meaning those configurations are prioritized.  
$\epsilon_\mathrm{parity}$, on the other hand, is a \emph{relative} error on the learned ground-state amplitudes, thus revealing this priority. 
An interesting direction for future work would be to investigate how our results change if we use a loss function that takes into account the relative errors of learned wavefunction amplitudes instead of their absolute errors.

\bibliographystyle{apsrev4-2}
\bibliography{main}

\end{document}